\documentclass[letterpaper,11pt]{article}

\usepackage[utf8]{inputenc}

\usepackage[top=3.25cm, bottom=3cm, left=3.5cm, right=3.5cm,           heightrounded,marginparwidth=1.5cm, marginparsep=1cm]{geometry} 

\usepackage[shortlabels]{enumitem} 

\usepackage[square,numbers,merge,comma,sort&compress]{natbib} 
\makeatletter
\def\NAT@spacechar{\,}  
\makeatother

\usepackage{amsmath,amssymb,amsfonts,amsthm,amsbsy}
\usepackage{mathtools} 

\usepackage{fnpct}
\setfnpct{dont-mess-around} 

\usepackage{slashed,cancel,latexsym}
\usepackage{old-arrows,comment,relsize,setspace,moresize,accents}
\usepackage{changepage,etoolbox}
\usepackage{epsfig}
\usepackage{mathrsfs,calligra,aurical,bold-extra} 
\usepackage{calc,float,appendix}
\usepackage{bigints,xargs,extarrows}
\usepackage{empheq,soul} 
\usepackage{upgreek}
\usepackage[all]{xy}
\usepackage[normalem]{ulem}

\usepackage[svgnames]{xcolor}  
\definecolor{Blue}{rgb}{0.0, 0.0, 0.37}
\definecolor{Green}{rgb}{0.05, 0.45, 0.25}
\xdefinecolor{dogwoodrose}{rgb}{0.8, 0.1, 0.55}
\xdefinecolor{RRed}{rgb}{0.7, 0.1, 0.4}

\usepackage{tikz}
\usetikzlibrary{scopes,decorations.pathmorphing,patterns,calc,arrows,shapes.geometric,shapes.arrows,decorations.markings,plotmarks}
\usepackage{pgfplots}

\usepackage[blocks]{authblk}  

\usepackage{graphicx} 
\graphicspath{{Figures/}} 

\usepackage[labelsep=colon]{caption}  
\usepackage[labelsep=colon,aboveskip=10pt,belowskip=10pt]{subcaption}
\DeclareCaptionSubType * [roman]{table}
\usepackage{titlesec}
\titleformat{\section}{\normalfont\fontsize{12.5}{12}\bfseries}{\thesection}{0.5em}{}
\titleformat{\subsection}{\normalfont\fontsize{10.5}{10}\bfseries}{\thesubsection}{0.5em}{}
\titleformat{\subsubsection}{\normalfont\normalsize\bfseries}{\thesubsubsection}{0.5em}{}

\titlespacing*{\section}{0pt}%
                {4ex plus 1ex minus .5ex}{1.75ex plus .25ex minus .25ex}
\titlespacing*{\subsection}{0pt}%
                {3.5ex plus 1ex minus .5ex}{1.25ex plus .2ex minus .2ex}
\titlespacing*{\subsubsection}{0pt}%
                {2.5ex plus 0.75ex minus .2ex}{0.75ex plus .15ex minus .15ex}
\titlespacing*{\paragraph}{0pt}%
                {1.85ex plus 0.5ex minus .15ex}{1em}

\usepackage{titletoc}
\addtocontents{toc}{\addvspace{-0.75em}}  
\usepackage{bm}  
\usepackage{dsfont}  

\DeclareMathAlphabet{\mathpzc}{OT1}{pzc}{m}{it}
\DeclareMathAlphabet{\mathcal}{OMS}{cmsy}{m}{n}
\DeclareSymbolFontAlphabet{\Scr}{rsfs}
\DeclareMathAlphabet{\mathbold}{U}{BOONDOX-ds}{m}{n}
\SetMathAlphabet{\mathbold}{bold}{U}{BOONDOX-ds}{b}{n}
\DeclareMathAlphabet{\mathcalboondox}{U}{BOONDOX-calo}{m}{n}
\SetMathAlphabet{\mathcalboondox}{bold}{U}{BOONDOX-calo}{b}{n}
\DeclareMathAlphabet{\mathbcalboondox}{U}{BOONDOX-calo}{b}{n}

\DeclareFontFamily{U}{matha}{\hyphenchar\font45}
\DeclareFontShape{U}{matha}{m}{n}{ <5> <6> <7> <8> <9> <10> gen * matha
                    <10.95> matha10 <12> <14.4> <17.28> <20.74> <24.88> matha12}{}
\DeclareSymbolFont{matha}{U}{matha}{m}{n}
\DeclareMathSymbol{\varleftarrow}{3}{matha}{"D0}
\DeclareMathSymbol{\varrightarrow}{3}{matha}{"D1}
\DeclareMathSymbol{\simeq}{3}{matha}{"14}
\DeclareMathSymbol{\sim}{3}{matha}{"12}
\DeclareMathSymbol{\ll}{3}{matha}{"21}
\DeclareMathSymbol{\gtrsim}{3}{matha}{"C1}
\DeclareMathSymbol{\lesssim}{3}{matha}{"C0}
\DeclareMathSymbol{\wedge}{2}{matha}{"5E}

\newcommand\linkcol{RRed}

\usepackage[breaklinks=true,backref=page]{hyperref}
\hypersetup{
    bookmarks=true,         
    bookmarksnumbered=true,
    pdftoolbar=true,        
    pdfmenubar=true,        
    pdffitwindow=false,     
    pdfauthor={},     
    pdfsubject={},   
    pdfcreator={},   
    pdfproducer={},  
    pdfpagemode={UseNone},
    pdfstartview={FitH},
    colorlinks=true,
    plainpages,
    linktoc=page,
    citecolor=blue,
    filecolor=black,
    linkcolor=\linkcol,
    urlcolor=Green,
}

\newif\ifbackrefshowonlyfirst
\backrefshowonlyfirsttrue
\makeatletter
\let\BR@direct@old@hyper@natlinkstart\hyper@natlinkstart
\renewcommand*{\hyper@natlinkstart}{\phantomsection\BR@direct@old@hyper@natlinkstart}
\let\BR@direct@oldBR@citex\BR@citex
\renewcommand*{\BR@citex}{\phantomsection\BR@direct@oldBR@citex}%
\long\def\hyper@page@BR@direct@ref#1#2#3{\hyperlink{#3}{#1}}
\ifx\backrefxxx\hyper@page@backref
    \let\backrefxxx\hyper@page@BR@direct@ref
    \ifbackrefshowonlyfirst
    \fi
\else
    \ifbackrefshowonlyfirst
    \fi
\fi
\RequirePackage{etoolbox}
\patchcmd{\Hy@backout}{Doc-Start}{\@currentHref}{}{\errmessage{I can't seem to patch backref}}
\makeatother

\renewcommand*{\backref}[1]{}
\renewcommand*{\backrefalt}[4]{%
\ifcase #1 %
\relax
\or
~{\small [\textsc{p.~\fns{\!#2}}]}
\else
~{\small [\textsc{p.~\fns{\!#2}}]}%
\fi}

\usepackage{footnotebackref}
\usepackage{hypernat} 

\usepackage{cleveref} 

\makeatletter
\normalsize
\makeatother

\makeatletter
\g@addto@macro\bfseries{\boldmath}
\makeatother

\def\+{~+~}
\def\-{~-~}
\def\={\:=\:}
\newcommand\fns{\footnotesize}

\newcommandx\Hodge[1][1=4,usedefault]{{}^{\star_{#1}}}
\newcommand\AdS{\mathrm{AdS}}
\newcommand\SO{\mathrm{SO}}

\newcommandx{\overbar}[1]{\mkern
1.5mu\overline{\mkern-2.0mu#1\mkern-2.0mu}\mkern 1.5mu}
\newcommandx{\overbarcal}[1]{\mkern                   6.0mu\overline{\mkern-5.5mu#1\mkern-1.0mu}\mkern 1.5mu}

\newcommand{\chiR}{\chi_{{}_{\text{R}}}}
\newcommand{\chiI}{\chi_{{}_{\text{I}}}}

\newcommandx{\eM}[1][1=A,usedefault]{\epsilon_{{}_{(\text{M})}}^{#1}}

\title{%
\vspace{-1em}
\centering\boldmath\Large\bfseries 
Supersymmetric smooth distributions of M2-branes as AdS solitons 
\bigskip\bigskip}

\author[1,2]{A.\ Anabal\'{o}n\footnote{\href{anabalo@gmail.com}{anabalo@gmail.com}}}

\author[3]{D.\ Astefanesei\footnote{\href{mailto:dumitru.astefanesei@pucv.cl}{dumitru.astefanesei@pucv.cl}}}

\author[4]{A.\ Gallerati \footnote{\href{mailto:antonio.gallerati@polito.it}{antonio.gallerati@polito.it}}}

\author[1]{J.\ Oliva\footnote{\href{julioolivazapata@gmail.com}{julioolivazapata@gmail.com}}}

\affil[1]{
Departamento de Física, Universidad de Concepción,
Casilla 160-C, Concepción, Chile.
\medskip
}

\affil[2]{
Instituto de F\'isica Te\'orica, Universidade Estadual Paulista,
R.\ Dr.\ Bento T.\ Ferraz 271, Bl.\ II, Sao Paulo 01140-070, Brazil
\medskip
}

\affil[3]{
Pontificia Universidad Católica de Valparaíso, Instituto de Física,
Av. Brasil 2950, Valparaíso, Chile%
\medskip
}

\affil[4]{
{Politecnico di Torino, Dipartimento di Scienza Applicata e Tecnologia, corso Duca degli Abruzzi 24, 10129 Torino, Italy}%
\medskip
}

\date{}

\begin{document}

\maketitle

\begin{abstract}
\noindent
We show that the singularities of certain distributions of M2-branes can be removed by adding a Wilson loop around a compact direction along the brane. The holographic coordinate naturally ends up before the singularity and the resulting spacetime provides an AdS soliton configuration, albeit possibly supersymmetric. We study the phase space diagram of these M2-branes distributions endowed with Wilson loops and show that different distributions are interconnected, providing a rich structure of phase transitions at zero temperature. 
\end{abstract}

\newpage

\tableofcontents
\newpage


\section{Introduction and discussion}
One of the most important open problems in theoretical particle physics is the description of confinement in non-abelian gauge theories. When an holographic description of the quantum field theory exists \cite{Maldacena:1997re} it was pointed out that this confined phase might have a very simple description in AdS \cite{Witten:1998zw}. A Lorentzian section of this everywhere-regular confining metric has negative energy and is perturbatively stable: the configuration is known as the AdS soliton \cite{Horowitz:1998ha}.\par
Perturbative stability is non-trivial, since the AdS soliton entails the existence of fermions with antiperiodic boundary conditions on a spacelike $\mathbb{S}^1$, which in asymptotically flat spaces gives rise to instabilities associated to the production of bubbles of nothing \cite{Witten:1981gj}. This, however, is not the case in AdS. Moreover, one can ensure a non-perturbative stability, due the existence of a BPS bound, when the AdS soliton is endowed with a magnetic flux \cite{Anabalon:2021tua, Anabalon:2023oge}. This opened the possibility to check whether these AdS solitons have more sophisticated supersymmetric configurations. In this regard, some of us recently studied a framework with a dilaton scalar running in the T${}^3$ truncation of the gauged maximal $\mathcal{N}=8$ supergravity \cite{Anabalon:2022aig}. This kind of construction has also been generalized to 10 dimensions, as new models of holographic confinement \cite{Canfora:2021nca, Nunez:2023nnl, Nunez:2023xgl,Fatemiabhari:2024aua}. The soliton/black hole phase transition has also been discussed in some of these cases \cite{Anabalon:2022ksf,Quijada:2023fkc,Durgut:2023rmu}.\par
In this paper we provide a further generalization of these AdS soliton-like geometries, and advance their understanding as we show that they have a nice interpretation as smoothing singular distributions of M2-branes in eleven dimensions. The latter were found and discussed in detail in \cite{Cvetic:1999xx}. Our results encompass some of these brane distributions and allow to construct their phase diagrams. We shall also see how the same distribution is associated to inequivalent spacetimes, and study how the latter can coexist for different values of the boundary conditions on the gauge fields. Indeed, the boundary value of the gauge fields can be interpreted as giving vacuum expectation values (vevs) to currents in a dual field theory.\par
In the following, we focus on a sector of the $D=4$, gauged $\mathcal{N}=8$ supergravity theory, breaking the isometry of the seven-sphere $\mathbb{S}^7$ from $\mathrm{SO}(8)$ to $\mathrm{SO}(4)\times\mathrm{SO}(4)$ for vanishing gauge fields. Hence, it is possible to capture only certain M2-branes distributions, namely those that have the same isometry. The construction is quite similar to what we found in \cite{Anabalon:2022aig}. In this respect, we believe that the connection between brane distributions and regular AdS solitons points to the possibility that various singular brane distributions can be smoothed by giving a vev to a current on the QFT side.\par
The structure of the paper is as follows. In Sect.\ \ref{sec:model}, we present a consistent truncation of the $\mathrm{SO}(8)$-gauged maximal supergravity framework, its oxidation to eleven dimensions and the associated BPS conditions. In Sect.\ \ref{sec:branes} we reproduce the brane distributions of \cite{Cvetic:1999xx}, specialized to the chosen truncation. We also present the change of coordinates that connect the $\mathbb{S}^7$ compactification yielding the STU model with the one that is better adapted to describe the brane distribution. We also introduce a change of coordinates to suitably describe the maximal analytic extension of the M2-brane spacetimes with $\mathrm{SO}(4)\times\mathrm{SO}(4)$ isometry. It therefore becomes clear that, for every value of the parameters of the distribution, there are two inequivalent and disconnected spacetimes, both of them asymptotically $\AdS_4\times \mathbb{S}^7$ and singular. In Sect.\ \ref{sec:solutions} we construct new spacetime configurations that generalize these distributions and yield a vev for the holographic current. The supersymmetric limit of these new soliton solutions are discussed in Sect.\ \ref{sec:susy}, and contrasted to the supersymmetric case in the absence of running scalars. Finally, in Sect.\ \ref{sec:phase} we compute the Euclidean action and energy of the solutions, carefully investigating their phase diagrams.

\section{The model}\label{sec:model}
We are interested in studying the dilatonic sector of the STU\ model of the maximal
$\mathrm{SO}(8)$-gauged, $\mathcal{N}=8$ supergravity with action
\begin{equation} 
\mathcal{S}=\frac{1}{2\,\kappa}\int\! d^{4}x\:\sqrt{-g}\left(R-\sum_{i=1}^{3} \frac{\left(\partial\Phi_{i}\right)  ^{2}}{2}+\frac{2}{L^{2}}\,\cosh\left(\Phi_{i}\right)
-\frac{1}{4}\:\sum_{i=1}^{4}X_{i}^{-2}\bar{F}_{i}^{2}\right)\:,
\label{eq:LagSTU}
\end{equation}
where $\bar{F}_{i}$ are two forms, related with gauge fields in the standard way, and we have
\begin{align}
\label{eq:11D_F_X_P}
\bar{F}_{i}=d\bar{A}_{i}\,,
\qquad
X_{i}=e^{-\frac{1}{2}\vec{a}_{i}\cdot
\vec{\Phi}}\,,
\qquad
\vec{\Phi}=\left(\Phi_{1},\Phi_{2},\Phi_{3}\right)\,,
\end{align}
with
\begin{equation}
\vec{a}_{1}=\left(1,1,1\right),
\qquad
\vec{a}_{2}=\left(1,-1,-1\right),
\qquad
\vec{a}_{3}=\left(-1,1,-1\right),
\qquad
\vec{a}_{4}=\left(-1,-1,1\right).\;
\end{equation}
We consider purely magnetic solutions, for which it is
consistent to truncate the axions to zero. The Lagrangian \eqref{eq:LagSTU} can be
obtained from the compactification of eleven dimensional supergravity over the
seven sphere with the ansatz \cite{Cvetic:1999xp}
\begin{align}
ds_{11}^{2}  =&\;\tilde{\Delta}^{2/3}\,ds_{4}^{2}+4\,L^{2}\tilde{\Delta}^{-1/3}\:\sum_{i=1}^{4}X_{i}^{-1}\left(d\mu_{i}^{2}+\mu_{i}^{2}\left(d\varphi_{i}+\frac{1}{2L}\bar{A}_{i}\right)^{2}\right)\:,
\\[1ex]
F =&-\frac{1}{L}\:\epsilon_{4}\:\sum_{i=1}^{4}\left(X_{i}^{2}\mu_{i}^{2}-\tilde{\Delta}\,X_{i}\right)  +L\,X_{i}^{-1}\:\Hodge dX_{i}\wedge d\mu_{i}^{2}
\nonumber\\
&-4\,L^{2}\:\sum_{i}X_{i}^{-2}\mu_{i}\,d\mu_{i}\wedge\left(d\varphi_{i}+\frac{1}{2\,L}\bar{A}_{i}\right)  \wedge\Hodge\bar{F}_{i}\:,
\end{align}
where $\epsilon_{4}$ is the volume form of the four-dimensional metric $ds_{4}^{2}$, $F$ is the four-form field strength and $\Hodge$ is the Hodge dual operator. The four independent rotations on $\mathbb{S}^{7}$ are parameterized by the $\varphi_{i}$, which are $2\pi$-periodic angular coordinates. We also have
\begin{equation}
\tilde{\Delta}=\sum_{i=1}^{4}X_{i}\,\mu_{i}^2\,,
\qquad
\sum_{i=1}^{4}\mu_{i}^2=1\,,
\end{equation}
where the latter $\mu_i$ can be parameterised
in terms of angles on the 3-sphere as 
\begin{equation}
\mu_1=\sin\vartheta\,,\quad
\mu_2=\cos\vartheta\,\sin\psi\,,\quad
\mu_3=\cos\vartheta\,\cos\psi\,\sin\xi\,,\quad
\mu_4=\cos\vartheta\,\cos\psi\,\cos\xi\,.
\end{equation}
We are interested in considering the higher-dimensional interpretation of some of our solutions
using this uplift. In particular, we shall work with a simplified, consistent truncation of the theory, where
\begin{equation}
\Phi_1=\Phi_3=0\,,\quad\;
\Phi_2=\sqrt{2}\,\phi\,,\quad\;
\Bar{F}^{1}=\Bar{F}^{3}=F^{1}\,,\quad\;
\Bar{F}^{2}=\Bar{F}^{4}=F^{2}\,,
\end{equation}
and we then obtain an action of the form:
\begin{equation}
\mathcal{S}=\frac{1}{\kappa}\int\! d^{4}x\:\,\sqrt{-g}\left(
    \frac{R}{2}-\frac{1}{2}\left(\partial\phi\right)^{2}
     -\frac{1}{4}\,e^{\sqrt{2}\,\phi}\left(F^{1}\right)^{2}
    -\frac{1}{4}e^{-\sqrt{2}\,\phi}\left(F^{2}\right)^{2}+\frac{1}{L^{2}}\left(2+\cosh\big(\sqrt{2}\,\phi\big)\right) \right),
\label{eq:Lag}
\end{equation}
where
\begin{equation}
F_{\mu\nu}^{\Lambda}=\partial_{\mu}A_{\nu}^{\Lambda}-\partial_{\nu}A_{\mu
}^{\Lambda}\,,
\qquad\;
\Lambda=1,2\,.   
\end{equation}
The above Lagrangian yields a consistent truncation to the dilaton $\phi$ only of the $\mathcal{N}=2$ \emph{minimal coupling} supergravity \cite{Luciani:1977hp}, provided the constraint
\begin{equation}
F^1\wedge F^1-e^{-2\sqrt{2}\,\phi}\,F^2\wedge F^2=0\:,
\label{eq:construnc}
\end{equation}
is satisfied. The minimal-coupling model defined by \eqref{eq:Lag} is a consistent truncation of the STU model, characterized by a scalar manifold of the form $\mathrm{SL}(2,\mathbb{R})/\SO(2)$.%
\footnote{%
The special geometry \cite{Gallerati:2016oyo,Lauria:2020rhc} of this smaller theory can be described in terms of a prepotential function $\mathcal{F}(\mathcal{X}^\Lambda)\:=-\frac{i}{4}\:\left(\mathcal{X}^0\right)\,\left(\mathcal{X}^1\right)$ and is selected among the class of theories discussed in \cite{Anabalon:2020pez} by choosing $n=1$, corresponding to $\nu=\infty$. The formulae obtained in the present paper can obtained by choosing the Fayet-Iliopoulos terms of the gauged $D=4$ supergravity as \,$\theta_M=\left(\rho^{-1}L^{-2},\,\rho/4,\,0,\,0\right)$, setting $\zeta=0$ in the dyonic $\nu=\infty$ model considered in \cite{Anabalon:2020pez}, having also suitably shifted the dilaton and rescaled the vector fields as described in the same reference.
}
%

\subsection{Supersymmetry}
The general formulae related to the supersymmetry transformations in  $\mathcal{N}=2$, $D=4$ supergravity with Fayet-Iliopoulos terms can be found in Appendix A of \cite{Anabalon:2022aig}, where we also define our spinor conventions.

\paragraph{Minimal-coupling model truncation.}
In order to see if the solutions preserve part of the supersymmetry, we consider the explicit form of the fermionic variations of the gauged supergravity theory. The latter, once adapted to the minimal-coupling model, read
\begin{align}
\delta\Psi^A_\mu\:=\;
  &\,\partial_\mu\epsilon^A+\frac{1}{4}\,{\omega_\mu}^{\!\!ab}\,\gamma_{ab}\,\epsilon^A+ \frac{1}{2\,L}\left(A^1_\mu+A^2_\mu\right)i\left(\sigma^2\right)^A{\!\!}_B\;\epsilon^B-
\nonumber\\[\jot]
  &-\frac{1}{8}\left(F^1_{\nu\rho}\,e^{\frac{1}{\sqrt{2}}\,\phi} +F^2_{\nu\rho}\,e^{-\frac{1}{\sqrt{2}}\,\phi}\right)\,\gamma^{\nu\rho}\,\gamma_\mu\,\varepsilon^{AB}\:\epsilon_B+
\nonumber\\[\jot]
  &+\frac{1}{2}\;\mathcal{W}\;\gamma_\mu\,\delta^{AB}\,\epsilon_B\:,
\\[3ex]
\delta\lambda^{A}=
  &-\gamma^\mu\,\partial_\mu\phi\:\epsilon^A  
  -\frac{1}{2\sqrt{2}}\left(-F^1_{\nu\rho}\,e^{\frac{1}{\sqrt{2}}\,\phi}
  +F^2_{\nu\rho}\,e^{-\frac{1}{\sqrt{2}}\,\phi}\right)\,\gamma^{\nu\rho}\,\varepsilon^{AB}\,\epsilon_B-
\nonumber\\
  &+\frac{1}{\sqrt{2}\,L}\,\left(e^{\frac{1}{\sqrt{2}}\,\phi}-e^{-\frac{1}{\sqrt{2}}\,\phi}\right)\,\delta^{AB}\,\epsilon_B\;,
\end{align}
where the superpotential explicitly reads
\begin{equation}
\mathcal{W}=\frac{e^{\frac{1}{\sqrt{2}}\,\phi}+e^{-\frac{1}{\sqrt{2}}\,\phi}}{2\,L}\:.
\end{equation}
We write the chiral spinors in terms of their real and imaginary components, using the spinor conventions specified in Appendix A of \cite{Anabalon:2022aig}:
\begin{equation}
\epsilon^{A}=\operatorname{Re}\epsilon^{A}+i\,\operatorname{Im}\epsilon^{A}\,,
\end{equation}
and introduce the complex spinors
\begin{equation}
\chiR
=\operatorname{Re}\epsilon^{1}+i\,\operatorname{Re}\epsilon^{2}\,,
\qquad\quad
\chiI=\operatorname{Im}\epsilon^{1}+i\,\operatorname{Im}
\epsilon^{2}\,.
\end{equation}
Each of the Killing spinor relations may be expressed as a first-order differential equation. Nevertheless, the two are not independent, and we can solve the spinor equations in just one of them ($\chiR$, for instance). Since in the chosen spinor basis
\begin{equation}
\epsilon_A=\left(\epsilon^A\right)^*\,,
\end{equation}
the Majorana spinors $\eM$ is expressed as
\begin{equation}
\eM=\epsilon_A+\epsilon^A=2\,\operatorname{Re}\epsilon^A\:.
\end{equation}
The action of $\gamma^5=i\,\gamma^0\gamma^1\gamma^2\gamma^3$ on the above Majorana $\eM$ is
\begin{equation}
\gamma^5\,\eM=\epsilon_A-\epsilon^A=-2\,i\,\operatorname{Im}\epsilon^A\,,
\end{equation}
so that
\begin{equation}
\operatorname{Im}\epsilon^A=i\,\gamma^5\,\operatorname{Re}\epsilon^A\:,
\end{equation}
giving
\begin{equation}
\chiI\equiv \operatorname{Im}\epsilon^1+i\,\operatorname{Im}\epsilon^2
=i\,\gamma^5\left(\operatorname{Re}\epsilon^1+i\,\operatorname{Re}\epsilon^2\right)=i\,\gamma^5\,\chiR\:,
\label{XIXR}
\end{equation}
providing the connection between the complex spinors.\par
%

The spinor components can be retrieved from a solution $\chiR$ of the Killing spinor equations as
\begin{equation}
\eM[1]=2\,\operatorname{Re}\chiR\,,
\qquad\quad
\eM[2]=2\,\operatorname{Im}\chiR\,,
\end{equation}
and
\begin{equation}
\begin{split}
\epsilon^1=\frac{\left(\mathds{1}-\gamma^5\right)}{2}\,\eM[1]=\left(\mathds{1}-\gamma^5\right)\,\operatorname{Re}\chiR\:,
\\[2.5ex]
\epsilon^2=\frac{\left(\mathds{1}-\gamma^5\right)}{2}\,\eM[2]=\left(\mathds{1}-\gamma^5\right)\,\operatorname{Im}\chiR\:.
\label{eq:epsfromchiR}
\end{split}
\end{equation}
The explicit equations for $\chiR$ have the form
\begin{align}
0\:=\:
  &\:\partial_\mu\chiR+\frac{1}{4}\,{\omega_\mu}^{\!\!ab}\,\gamma_{ab}\,\chiR - \frac{i}{2\,L}\,\left(A^1_\mu + A^2_\mu\right)\chiR+
\nonumber\\[\jot]
  &+\frac{i}{8}\,\left(F^1_{\nu\rho} \,e^{\frac{1}{\sqrt{2}}\,\phi}
  +F^2_{\nu\rho}\,e^{-\frac{1}{\sqrt{2}}\,\phi}\right)\gamma^{\nu\rho}\,\gamma_\mu\,\chiR+\frac{1}{2}\;\mathcal{W}\;\gamma_\mu\,\chiR\:,
\\[2.5ex]
0\:=\:
  &-\gamma^\mu\,\partial_\mu\phi\:\chiR
  +\frac{i}{2\sqrt{2}}\left(-F^1_{\nu\rho}\,e^{\frac{1}{\sqrt{2}}\,\phi}
  +F^2_{\nu\rho}\,e^{-\frac{1}{\sqrt{2}}\,\phi}\right)\gamma^{\nu\rho}\,\chiR+
\nonumber\\[0.5\jot]
  &+\frac{1}{\sqrt{2}\,L}\,\left(e^{\frac{1}{\sqrt{2}}\,\phi}-e^{-\frac{1}{\sqrt{2}}\,\phi}\right)\chiR\;.
\end{align}
To derive analogous conditions on $\chiI$ it is sufficient to multiply the above expressions by \,$i\,\gamma^5$ from the left.

\section{Maximal extension of a M2 brane distribution}\label{sec:branes}
Now we will take a singular distribution of M2-branes and show how it can be seen as a four dimensional solution of the theory \eqref{eq:Lag}. While the solution is known, the maximal extension we provide in this section is new.\par
Our starting point is the 11-dimensional M2 brane distribution of \cite{Cvetic:1999xx}:
\begin{equation}
\begin{split}
ds_{11}^2&=H^{-2/3}\left(-dt^2+dz^2+d\varphi^2\right)+H^{1/3}\,ds^2_8\:,
\\[1em]
F^{(4)}&= dt\wedge dz \wedge d\varphi\wedge dH^{-1}\:,
\end{split}
\end{equation}
where 
\begin{equation}
H=\frac{(2 L)^6}{\rho^6
\,\Delta_0}\,,\qquad\quad \Delta_0=(H_1 ...H_8)^{1/2}\,\sum_{i=1}^{8} \frac{y^2_i}{H_i}\, ,
\end{equation}
with
\begin{equation}
H_i=1+\frac{\ell_i^2}{\rho^2}
\end{equation}
and where $ds^2_8$ is just flat space in disguise,
\begin{equation}
ds^2_8=\frac{\Delta_0\,d\rho^2}{(H_1\cdots H_8)^{1/2}}+\rho^2\,\sum_{i=1}^{8} H_i\, dy^2_i\:,
\end{equation}
with the constraint %
\begin{equation}
\sum_{i=1}^{8}y_i^2=\text{constant}\:.    
\end{equation}
The unconstrained cartesian coordinates are
\begin{equation}
w^i=\rho\,\sqrt{H_i}\: y_i\:,
\end{equation}
and the metric is given by 
\begin{equation}
ds^2_8=\delta_{i j}\,dw^{i} dw^{j}\:. \end{equation}
In order to make contact with the STU model, we specialize to the case when
\begin{equation}
\ell_8=\ell_1\,,\;\quad \ell_7=\ell_2\,,\;\quad  \ell_6=\ell_3\,,\;\quad\ell_5=\ell_4\,,
\end{equation}
and we introduce the new coordinates
\begin{equation}
\begin{split}
y_1 = \mu_1\,\cos\varphi_1\,,\qquad
y_8 = \mu_1\,\sin\varphi_1\,, \\[\jot]
y_2 = \mu_2\,\cos\varphi_2\,, \qquad 
y_7 = \mu_2\,\sin\varphi_2\,, 
\\[\jot]
y_3 = \mu_3\,\cos\varphi_3\,,\qquad 
y_6 = \mu_3\,\sin\varphi_3 \,,
\\[\jot]
y_4 = \mu_4\,\cos\varphi_4\,, \qquad 
y_5 = \mu_4\,\sin\varphi_4 \,.   
\end{split}
\end{equation}
Using this new parametrization, we see that  
\begin{equation}
\Delta_0=(H_1\cdots H_4)^{3/4}\,\sum_{i=1}^{4}X_i \,\mu^2_i=(H_1\cdots H_4)^{3/4}\,\tilde{\Delta}\, ,\qquad\;
X_i=\frac{(H_1\cdots H_4)^{1/4}}{H_i}\,.\quad
\end{equation}
The eleven-dimensional M2-branes metric now reads
\begin{equation}\label{eq:A1}
\begin{split}
ds_{11}^2=\:&\tilde{\Delta}^{2/3}\:\frac{\rho^4 (H_1\cdots H_4)^{1/2}}{2^4 L^4}\left(-dt^2+dz^2+d\varphi^2+G(\rho)\,d\rho^2\right)+
\\[\jot]
&+4\,L^2\,\tilde{\Delta}^{-1/3}\,\sum_{i=1}^{4} X_i^{-1} \left(d\mu_i^2+\mu_i^2\,d\varphi_i^2\right)\,,
\end{split}
\end{equation}
with
\begin{equation}
G(\rho)=\frac{2^6 L^6}{\rho^6\, H_1\cdots H_4}\:.
\end{equation}
The distribution is read-off from the harmonic function $H$, which can be written as
\begin{equation}
H=(2L)^{6}\int \frac{\sigma(w')}{|w-w'|^6}\:d^8 w'\,,
\qquad\;
\sigma=\frac{2}{\pi^4 \,\ell_1^2\,\ell_2^2\,\ell_3^2\,\ell_4^2}\:\delta'\left(1-\rho^2\sum_{i=1}^{4}\frac{H_i\,\mu_i^2}{\ell_i^2}\right).
\end{equation}
Note that \eqref{eq:A1} indeed coincides with the sphere reduction of the previous section, whenever $\sum_{i=1}^{4}\mu_i^2=1$ and the gauge fields vanish. The truncation we are interested in is obtained setting $\ell_3=\ell_1$,  $\ell_4=\ell_2$. In this latter case, the distribution has support on an ellipsoid in terms of the Cartesian coordinates $w_i$:
\begin{equation}
\frac{w_1^2+w_3^2+w_6^2+w_8^2}{\ell_1^2}+\frac{w_2^2+w_4^2+w_5^2+w_7^2}{\ell_2^2}=1\:,
\end{equation}
which makes the isometry of the distribution manifestly $\SO(4)\times \SO(4)$ invariant. In terms of the intrinsic coordinates of the $\mathbb{S}^7$, this reads:
\begin{equation}\label{dist}
\rho^2\left(\frac{\mu_1^2+\mu_3^2}{\ell_1^2}+\frac{\mu_2^2+\mu_4^2}{\ell_2^2}\right)=0\:.
\end{equation}
The term in parenthesis of \eqref{dist} is manifestly positive definite on the $\mathbb{S}^7$. Hence, we learn that the support of the M2 is located at $\rho=0$. We shall see below that this is not a curvature singularity. Indeed, the curvature singularity is actually located at negative values of $\rho^2$. We shall cure the latter in the M2 brane distribution and see that this yields a form of an AdS soliton in four dimensions, with a possible supersymmetric limit as in \cite{Anabalon:2021smx}.\par
We find convenient to introduce the following change of coordinates \cite{Anabalon:2012ta,Anabalon:2019tcy}
\begin{equation}
x=\frac{H_2}{H_1}
\quad\to\quad \rho^2=\frac{\ell_2^2-\ell_1^2\,x}{x-1}\:,
\end{equation}
where the four dimensional metric (the part multiplying $\tilde{\Delta}^{2/3}$) in \eqref{eq:A1} reads
\begin{equation}
ds_4^2=\Upsilon(x)\left(-dt^2+dz^2+d\varphi^2+\frac{\eta^2}{x^2}\:dx^2\right),
\end{equation}
with
\begin{equation}
\Upsilon(x)=\frac{L^{2}\,x}{\eta^{2}\left(x-1\right)^{2}}\,,\qquad
\eta^2=\frac{16\,L^6}{\left(\ell_1^2-\ell_2^2\right)^2}
\end{equation}
and the four dimensional dilaton reading \,$\phi=-\dfrac{\ln(x)}{\sqrt{2}}$.\par\smallskip
We note that the asymptotic region is located at $x=1$, namely for $\rho^2=\pm \infty$. Hence, we find two different disconnected geometries, $0<x<1$ and $1<x<\infty$\,. In particular, one has
\begin{equation}
\begin{alignedat}{3}
&\ell_1^2<\ell_2^2\::
\quad\;\;
&&x\in(0,1)\:\to\: \rho^2\in\left(-\infty,-\ell_2^2\right),\quad\;
&&x\in(1,+\infty)\:\to\: \rho^2\in\left(-\ell_1^2,+\infty\right),\;
\\[1.5em]
&\ell_1^2>\ell_2^2\::
\quad
&&x\in(0,1)\:\to\: \rho^2\in\left(-\ell_2^2,+\infty\right),\quad
&&x\in(1,+\infty)\:\to\: \rho^2\in\left(-\infty,-\ell_1^2\right).
\end{alignedat}
\end{equation}
Then, in any case, the coordinate $x$ provides a maximal analytic extension of the manifold, unveiling its geometry in a transparent form.\par
It should also be clear that the singularities in the manifold are not related to the M2 brane distributions. They occur at \,$\rho^2=-\ell_1^2$\, and $\rho^2=-\ell_2^2$\,. In the following, we will study how to remove these singularities. We will also see that the two geometries are actually interconnected from the boundary point of view, as they are both smoothly connected in the phase space description of these distributions in terms of the vevs of a dual current.

\section{Hairy soliton solutions}
\label{sec:solutions}
The model \eqref{eq:Lag} admits soliton solutions generalizing the configurations of \cite{Anabalon:2019tcy} with the presence of a scalar field. These solutions can be obtained by analytic continuations
\begin{equation}
t\to i\,\varphi\,,
\qquad\quad
\varphi\to i\,t\,,
\qquad\quad
Q_\Lambda\to i\,Q_\Lambda\,,
\label{eq:Wick}
\end{equation}
of the black hole configurations of \cite{Anabalon:2012ta, Anabalon:2017yhv, Anabalon:2020pez}, whose charged planar solutions can be embedded in the STU model \cite{Cvetic:1999xp}.%
\footnote{%
The STU model \cite{Duff:1995sm,Behrndt:1996hu,Behrndt:1997ny} is a $\mathcal{N}=2$ supergravity coupled to 3 vector multiplets and characterized, in~a suitable symplectic frame, by~the prepotential
\,${\mathcal{F}_\textsc{stu}(\mathcal{X}^\Lambda)\:=-\frac{i}{4}\,\sqrt{\mathcal{X}^0\,\mathcal{X}^1\,\mathcal{X}^2\,\mathcal{X}^3}}$,\,
together with symmetric scalar manifold of the form \,${\mathscr{M}_\textsc{stu}=\left(\mathrm{SL}(2,\mathbb{R})/\mathrm{SO}(2)\right)^3}$ spanned by the three complex scalars \,$z^i=\mathcal{X}^i/\mathcal{X}^0$\, ($i=1,2,3$);\, this model is in turn a consistent truncation of the maximal $\mathcal{N}=8$ theory in four dimensions with $\mathrm{SO}(8)$ gauge group \cite{Duff:1999gh,Andrianopoli:2013kya,Andrianopoli:2013jra,Andrianopoli:2013ksa}.
}\par
In particular, if we consider the dyonic, charged planar black hole of \cite{Anabalon:2020pez} for $\nu=\infty$, a hairy soliton solution obtained through \eqref{eq:Wick} 
reads%
\footnote{Note that the coordinates $(z,t)$ of this section are related to the one of the previous section by a factor $L$.}
\begin{equation}
\begin{split}
e^{0} & =\sqrt{\Upsilon(x)}\,L\,dt,
\quad\;
e^{1}=\sqrt{\frac{\Upsilon(x)}{f(x)}}\:\frac{\eta}{x}\,dx,
\quad\;
e^{2}=\sqrt{\Upsilon(x)\,f(x)}\,d\varphi,
\quad\;
e^{3}=\sqrt{\Upsilon(x)}\,L\,dz,
\\[1ex]
\phi&=-\frac{\ln(x)}{\sqrt{2}},
\quad\;
A^{1}=-P_1\,z\,dt+Q_{1}(x-x_{0})\,d\varphi,
\quad\;
A^{2}=-P_2\,z\,dt-Q_{2}\left(\frac{1}{x}-\frac{1}{x_{0}}\right)d\varphi,
\end{split}
\end{equation}
with
\begin{align}\label{Fx}
\Upsilon(x)=\frac{L^{2}\,x}{\eta^{2}\left(x-1\right)^{2}}\:,
\qquad
f(x) =1-\frac{\eta^{2}(x-1)^{3}}{L^{2}\,x}\left(Q_{1}^{2}
-\frac{Q_{2}^{2}}{x}-\frac{\eta^2}{L^4}\,\frac{P_1^2}{x}+\frac{\eta^2}{L^4}\,P_2^2\right).
\end{align}
The quantities of the $D=11$ theory of Sect.\ \ref{sec:model} can be written for this solution as
\begin{equation}
\begin{split}
&\Phi_2=-\log x\,,
\qquad
\Phi_1=\Phi_3=0\,,
\\[2ex]
&\Bar{F}^{1}=\Bar{F}^{3}=Q_1\;dx\wedge d\varphi\,,
\qquad
\Bar{F}^{2}=\Bar{F}^{4}=Q_2\;x^{-2}\,dx\wedge d\varphi\,,
\\[2ex]
&X_i=\left(x^{\frac{1}{2}}\,,\:x^{-\frac{1}{2}}\,,\:x^{\frac{1}{2}}\,,\:x^{-\frac{1}{2}}\right),
\\[2ex]
&\tilde{\Delta}=x^{\frac{1}{2}}\,\sin^2\vartheta+x^{-\frac{1}{2}}\,\cos^2\vartheta\,\left(\sin^2\psi+\cos^2\psi\left(\cos^2\xi+x\,\sin^2\xi\right)\right)\, ,
\end{split}
\end{equation}
which shows that everything is regular for $x\neq 0$ and $x\neq \infty$ in the eleven dimensional geometry. The constraint \eqref{eq:construnc} is solved if 
\begin{equation}
P_1\,Q_1-P_2\,Q_2\=0\:
\end{equation}
This indeed coincide with the brane distribution of the previous Section when there are no gauge fields.

\paragraph{Asymptotic expansions.}
The metric's conformal boundary is at $x=1$, where the conformal factor $\Upsilon(x)$ features a pole of order two. The metric then describes two distinct spacetimes for $x\in(0,1)$ and $x\in(1,\infty)$, physically identified by different signs of the dilaton field.\par
Let us restrict to a soliton magnetically-charged configuration, $P_1=P_2=0$. The canonical form of an asymptotically AdS$_{4}$ spacetime can be obtained by introducing the change of variable
\begin{equation}
x=1\pm\left(  \frac{L^{2}}{\eta\,\rho}-\frac{L^{6}}{8\,\eta^{3}\,\rho^{3}}\right)
+\frac{L^{4}}{2\,\eta^{2}\,\rho^{2}}+O(\rho^{-3})\:,
\end{equation}
the sign determined by the choice of branch $x>1$ \,or\, $0<x<1$. This gives for the metric quantities
\begin{align}
\Upsilon(x)  &  =\frac{\rho^{2}}{L^{2}}+O(\rho^{-2})\:,
\\[\jot]
g_{\varphi\varphi}  &  =\Upsilon(x)\,f(x)=\frac{\rho^2}{L^2}-\frac{\mu}{\rho
}+O(\rho^{-2})\:,
\\[\jot]
\mu & =\pm\,\frac{L^{2}}{\eta}\left(Q_{1}^{2}-Q_{2}^{2}\right)\:,\label{eq:mu}
\end{align}
the latter $\mu$ being the energy parameter of the solution. The
dilaton field can be expanded as
\begin{align}
\phi=L^{2}\,\frac{\phi_{0}}{\rho}+L^{4}\,\frac{\phi_{1}}{\rho^{2}}+O(\rho^{-3})\:,
\end{align}
where
\begin{equation}
\phi_{0}   =\mp\frac{1}{\sqrt{2}\,\eta}\,,
\qquad
\phi_1=0\,,
\end{equation}
then giving only a non-trivial leading term $\phi_0$, which yields a vev in the QFT.

\paragraph{Dual theory.}
Our focus lies in soliton solutions, identified by the contraction of the $\varphi$-circle at point $x_0$ where
\begin{equation}
f(x_0)=0\:, 
\label{eq:fx0}
\end{equation}
the soliton configuration existing in the interval between $x_0$ and the boundary $x=1$.\par
Expanding the metric around $x_0$, together with the above condition \eqref{eq:fx0}, regularity requires defining a parameter $\Delta$ such that
\begin{equation}
\varphi\:\in\:[0,\,\Delta]\:,
\end{equation}
where $\Delta$ is given by
\begin{equation}
\Delta^{-1}
    =\left\vert \frac{1}{4\,\pi\,\eta}\:x\,\frac{df}{dx}\right\vert_{{}_{x=x_{0}}}
    =\:\left\vert \frac{\eta\left(x_0-1\right)^{2}}{4\,\pi\,L^{2}\,x_0^2}\Big(Q_1^2\,x_0\,(1+2\,x_0) -Q_{2}^{2}\,(2+x_0)\Big)\right\vert \:.
\end{equation}
Magnetically charged configurations feature net magnetic fluxes at infinity,
\begin{equation}
\label{eq:fluxes}
\begin{split}
\Phi_{\textsc{m}}^{1} & =\int F^{1}=\oint A^{1}=Q_{1}\,\Delta\left(1-x_{0}\right)\equiv2\pi
L\,\psi_{1},
\\[2ex]
\Phi_{\textsc{m}}^{2}  &  =\int F^{2}=\oint A^{2}=Q_{2}\,\Delta\left(-1+x_{0}^{-1}\right)\equiv2\pi
L\,\psi_{2}\:.
\end{split}
\end{equation}
%
The dilaton field generates a non-zero vev of an operator, in the dual theory, of conformal dimension one, that can be expressed in terms of the fluxes as
\begin{equation}
\label{eq:vev}
\left\langle\mathcal{O}\right\rangle=\phi_{0}=
  \mp\frac{\pi}{\sqrt{2}\,\Delta}\,x_0^{-1}\,\Big((1+2\,x_0)\,\psi_1^2-x_0\,(2+x_0)\,\psi_2^2\Big)\,.
\end{equation}
The dual energy momentum tensor is expressed in terms of the energy parameter as \cite{Myers:1999psa,Anabalon:2015xvl,Astefanesei:2018vga,Astefanesei:2021ryn}
\begin{equation}
\left\langle T_{tt}\right\rangle =-\frac{\mu}{2\,\kappa\, L^{2}}\,,
\qquad
\left\langle T_{zz}\right\rangle =\frac{\mu}{2\,\kappa\,L^{2}}\,,
\qquad
\left\langle T_{\varphi\varphi}\right\rangle =-\frac{\mu}{\kappa\,L^{2}}\,.
\end{equation}
Finally, from the presence of the gauge fields originates a vev for the currents at the boundary
\cite{Marolf:2006nd}
\begin{align}
\left\langle J^{\nu}_{1}\right\rangle  &  =\frac{\delta\mathcal{S}}{\delta A^{1}_{\nu}}=-\frac{1}{\kappa}\,N_{\mu}\,e^{\sqrt{2}\,\phi}\,F^{1\,\mu\nu}\,\sqrt{\left\vert h\right\vert}=-\frac{Q_{1}}{\eta\,\kappa}\,\delta_{\varphi}^{\nu}\:,\label{GC1}
\\[2ex]
\left\langle J^{\nu}_{2}\right\rangle  &  =\frac{\delta\mathcal{S}}{\delta A^{2}_{\nu}}=-\frac
{1}{\kappa}\,N_{\mu}\,e^{-\sqrt{2}\,\phi}\,F^{2\,\mu\nu}\,\sqrt{\left\vert
h\right\vert}=-\frac{Q_{2}}{\eta\,\kappa}\,\delta_{\varphi}^{\nu}\:, \label{GC2}
\end{align}
where $N_{\mu}$ is the outward pointing normal to the boundary metric
$h_{\mu\nu}=g_{\mu\nu}-N_{\mu}N_{\nu}$.

\subsection{Existence of solitons}
As seen from a bulk perspective, $Q_1$, $Q_2$, and $\eta$ parametrize the solutions. Nonetheless, from a boundary perspective, it makes more sense to parameterize solutions in terms of the boundary data we hold fixed.\par 
We can take into account fixed charges (or better, fixed currents) holding fixed $Q_1/\eta$, $Q_2/\eta$, and the period $\Delta$, or fixed fluxes (Wilson loops), holding fixed $\psi_1$, $\psi_2$, and the period $\Delta$.

\begin{figure}[H]
\centering
\includegraphics[width=0.8\textwidth,keepaspectratio]{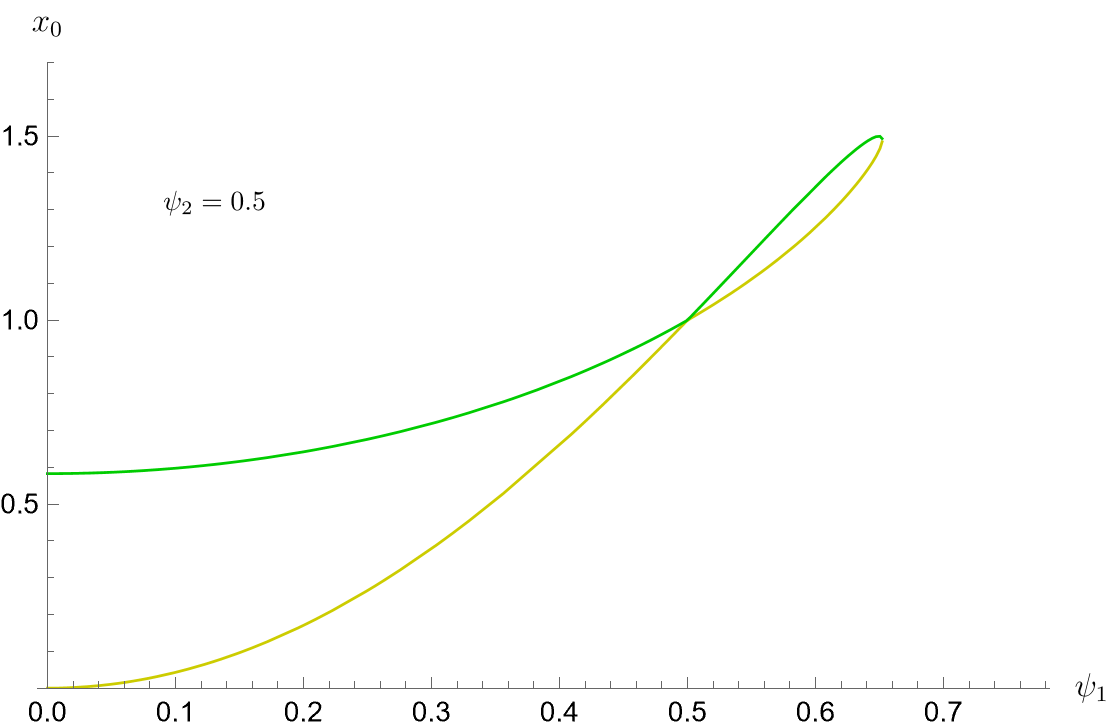}
\caption{Location of $x_0$ as a function of $\psi_1$, for fixed $\psi_2=0.5$\,. We find that, for each value of $\psi_1$ identifying a configuration, there exist two solutions for $x_0$, possibly coalescing.}%
\label{fig:sols_fixed_fluxes}
\end{figure}

\paragraph{Fixed fluxes.} 
Let us consider soliton configuration in terms of the boundary quantities $\psi_{1}, \psi_{2}$ and $\Delta$. We recast the bulk parameters as
\begin{equation} \label{Qfrompsi}
\begin{split}
&Q_{1}=\frac{2\pi\,L\,\psi_{1}}{\Delta\,(1-x_{0})}\,,
\qquad Q_{2}=\frac{2\pi\,L\,\psi_{2}}{\Delta\left(x_0^{-1}-1\right)}\,,\qquad\qquad
\\[2ex]
&\eta=\frac{x_{0}\,\Delta}{\pi\left\vert \psi_{1}^{2}\left(1+2\,x_{0}\right)-\psi_{2}^{2}\:x_0\,\left(2+x_0\right)\right\vert }\,.
\end{split}
\end{equation}
Substituting into $f(x_0)$, we get
\begin{equation} \label{fxp}
f(x_0)  = 1 + \frac{4\,x_0\,(1-x_0)\left(\psi_1^2-x_0\,\psi_2^2\right)}{\left((1+2\,x_0)\,\psi_1^2 -x_0\,(2+x_0)\,\psi_2^2\right)^2}\;. 
\end{equation}
Then, similarly to the Einstein-Maxwell configurations \cite{Kastor:2020wsm, Anabalon:2021tua}, there exists a limited range of parameters characterized by two different solitonic solutions, coalescing at the range extremum, see Fig.\ \ref{fig:sols_fixed_fluxes}.

\begin{figure}[H]
\centering
\includegraphics[width=0.9\textwidth,keepaspectratio]{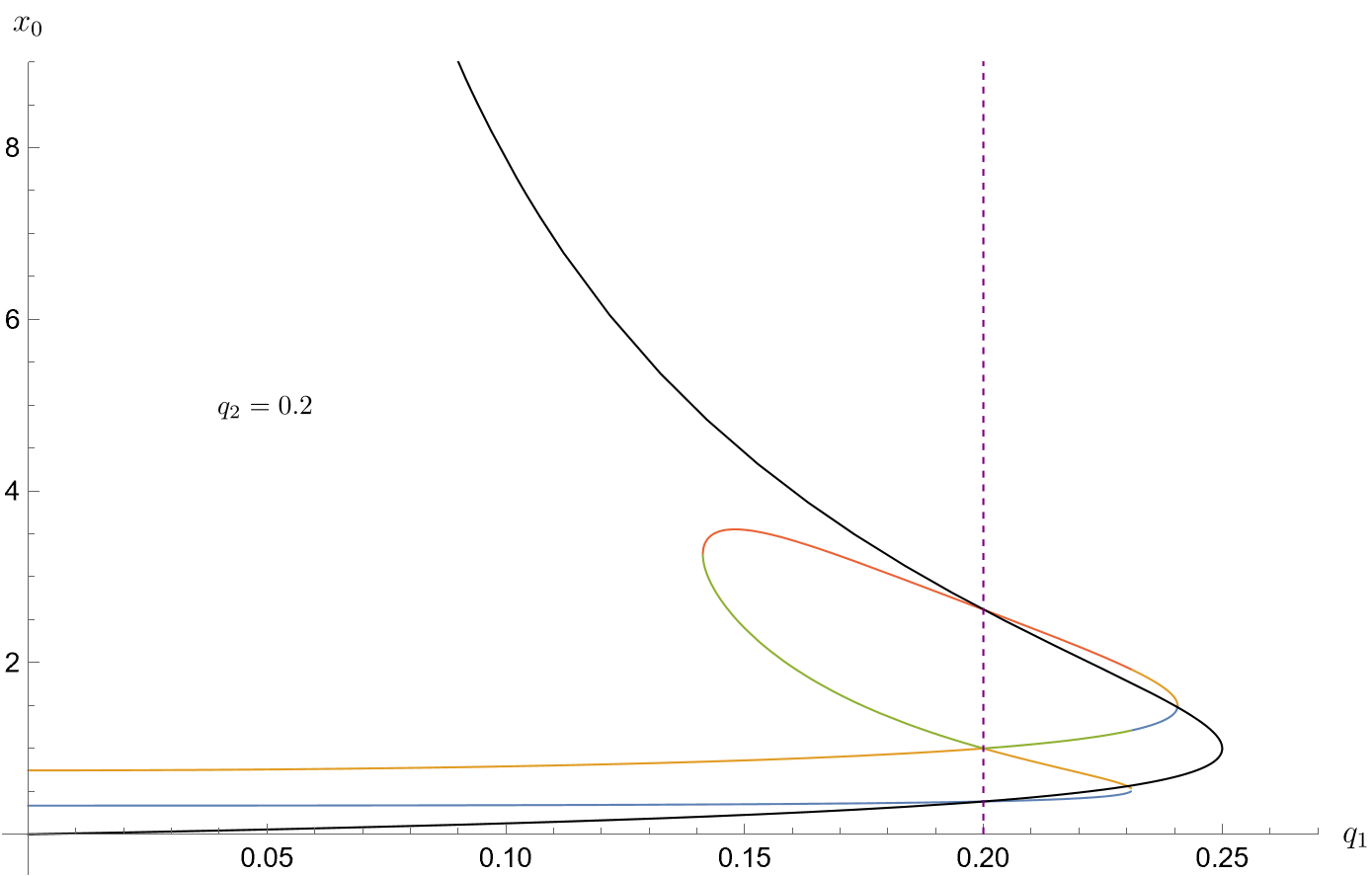}
\caption{The $x_0$ solutions for the roots of \eqref{eq:fx0_fixch} as functions of the rescaled charge $q_1$. The black line shows the location of the supersymmetric solitons, with the susy condition ${q_2=q_1}$ (see Sect.\ \ref{sec:susy}). The other coloured lines represent the different roots of \eqref{eq:fx0_fixch} for fixed ${q_2=0.2}$\,. The vertical dotted purple line corresponds to the value of $q_1$ satisfiying the susy condition $q_1=q_2=0.2$. As expected, this intersects the coloured solution lines where they intersect the black line. The additional intersection at $x_0=1$ corresponds to non-susy solutions with vanishing scalar (see following Subsect.\ \ref{subsec:susy_fixch}).
}%
\label{fig:sols_fixed_charges_1}
\end{figure}

\paragraph{Fixed charges.}
Let us now consider the boundary fixed charge framework. We introduce the rescaling 
\begin{equation}
q_{1,2}\equiv\frac{\Delta^{2}}{4\pi^{2}L}\,\frac{Q_{1,2}}{\eta}\:.
\label{eq:q1q2}
\end{equation}
If we rewrite the metric in terms of the rescaled charges and impose regularity around $x_0$, we obtain for $\eta$ the expression
\begin{equation}
\eta =\frac{\Delta}{4\pi}\:\frac{ q_{2}^{2}\,(2+x_0)-q_{1}^{2}\,x_0\,(1+2\,x_0) }{(-1+x_{0})\left(q_{2}^{2}-q_{1}^{2}\,x_0\right)}\;.
\label{eq:eta_fixch}
\end{equation}
The location of soliton configurations now come from the solutions of the equation
\begin{equation}
f(x_{0})=1-\frac{\left(  q_{2}^{2}\,(2+x_0)-q_{1}^{2}\,x_0\,(1+2\,x_0)\right)^{4}}{16\,x_0^{2}\,(-1+x_0)\left(q_1^2\,x_{0}-q_{2}^{2}\right)^3}\=0\;.
\label{eq:fx0_fixch}
\end{equation}
In Fig.\  \ref{fig:sols_fixed_charges_1} we show how, for every value of the rescaled charges \eqref{eq:q1q2}, from zero to four solutions can be found.%
\footnote{%
The situation is very different than for pure Einstein-Maxwell-AdS system, where there are only two solitons for each value of the charge \cite{Anabalon:2021tua}.
} 
The supersymmetric scenario (see Sect.\ \ref{sec:susy}) turns out to be simpler, featuring two superymmetric solitons for the same boundary conditions, with the same energy and free energy.%

\subsection{Relation to earlier solutions}
\label{subsec:earlier}
Soliton solutions in a model with a single gauge field were discovered in \cite{Anabalon:2021tua}, exploiting a compactification from the $D=11$ supergravity setup with \,$\bar A_i = \frac{1}{2}\,A$  \,($i = 1, \ldots 4$)\, and \,$\Phi_a =0$ \,($a=1,2,3$). The hairy solutions we discuss in this paper should then reduce to the scalar-free model \cite{Anabalon:2021tua} for \,$A^1 = A^2 = \frac{1}{2}\,A$\, and \,$\phi =0$.\par\smallskip
Let us know construct the explicit connection between the two models. We start with the change of coordinates
\begin{equation}
x = 1 - \frac{\alpha}{y}\:,
\end{equation}
setting the boundary at $y \to \infty$ and implying for the $x_0$ point location $x_0=1-\frac{\alpha}{y_0}$. 
The solution can be then rewritten as
\begin{equation}
\begin{split}
&e^{0}=\sqrt{\Upsilon(y)}\,dt,\quad
e^{1}=\sqrt{\frac{\Upsilon(y)}{f(y)}}\,\frac{\alpha\,\eta}{y\,(y-\alpha)}\,dy,\quad
e^{2}=\sqrt{\Upsilon(y)\,f(y)}\,d\varphi,\quad
e^{3} =\sqrt{\Upsilon(y)}\,dz,
\\[1ex]
&\phi=-\frac{1}{\sqrt{2}}\,\ln\left(1-\frac{\alpha}{y}\right),
\quad\;
A^{1}=Q_1\,\alpha\left(-\frac{1}{y} +\frac{1}{y_0} \right)\,d\varphi\,,
\quad\;
A^{2}=Q_2\,\alpha\,\frac{y-y_0}{(y-\alpha)\,(y_0-\alpha)}\,d\varphi,
\end{split}    
\end{equation}
with
\begin{equation}
\Upsilon(y)  =\frac{L^{2}\,y\,(y-\alpha)}{\alpha^2\,\eta^{2}}\,,
\qquad\quad
f(y) =1-\frac{\eta^{2}\,\alpha^3 \left(-Q_{1}^{2}
\,(y-\alpha)+Q_{2}^{2}\,y\right)}{L^{2}\,y^2\,(y-\alpha)^2}\,.\qquad
\end{equation}
To make contact with the model \cite{Anabalon:2021tua}, we have to impose $\phi=0$ while also obtaining the same functional form for non-vanishing gauge fields $A^1$ and $A^2$. In order to achieve this result in a well-defined configuration, we consider the simultaneous scalings
\begin{equation}
\alpha \to 0\,,\qquad
\eta \to \frac{L^2}{\alpha}\,,
\qquad
Q_{1,2}\to\eta\,\tilde Q_{1,2}\,,
\qquad
\tilde Q_1^2-\tilde Q_2^2\to\frac{\mu}{\eta\,L^2}\,,
\label{eq:limitsEM}
\end{equation}
resulting in a finite well-behaving limit for the solution and a standard form for the conformal factor ${\Upsilon \to \tfrac{y^2}{L^2}}$. \par
Explicitly, the above limit gives
\begin{equation}
\begin{split}
&e^{0} =\frac{y}{L}\,dt\,,\qquad
e^{1}  = \frac{dy}{\sqrt{f_0(y)}}\,,\qquad
e^{2} =\sqrt{f_0(y)}\,d\varphi\,,\qquad
e^{3} =\frac{y}{L}\,dz\,,\qquad
\\[1.5ex]
&\phi=0\,,
\qquad 
A^1 = A^2 
= L^2\,\tilde Q_1 \left( -\frac{1}{y} + \frac{1}{y_0} \right) d \varphi\,,
\end{split}
\end{equation}
with
\begin{equation}
f_0(y)= \frac{y^2}{L^2}\,f(y)= \frac{y^2}{L^2} - \frac{\mu}{y} - \frac{L^4\,\tilde{Q}_1^2}{y^2}\:, 
\end{equation}
which finally corresponds to the scalar free, single gauge field solution \cite{Anabalon:2021tua} for $Q=L^2\,\tilde Q_1$. 
\par\smallskip

\section{Supersymmetric solutions}
\label{sec:susy}
The soliton configuration preserve some part of the supersymmetry if
\begin{equation}
Q_{1}=Q_{2}
\quad\Longrightarrow\quad
\mu=0\;.
\label{eq:susyQ}
\end{equation} 
It is possible to achieve the aforementioned condition by solving the Killing spinor equations. We also remark that, as expected, supersymmetry leads to the vanishing of the energy parameter $\mu$, see \eqref{eq:mu},  while the reverse is not true (the solution is not supersymmetric for $Q_1 =- Q_{2}$).\par
Applying the susy condition \eqref{eq:susyQ} in the $f(x)$ function gives
\begin{equation}
f(x)=1-\frac{Q_1^2\:\eta^2\,(-1+x)^4}{L^{2}\,x^2}\:,
\end{equation}
and the Killing spinors can be explicitly written as
\begin{alignat}{2}
\chi_{{}_{\text{R}_{(1)}}} &  =e^{i\,\omega\,\varphi}\left(
\begin{array}{@{}c@{}}
\alpha_{-}(x) \\
0\\
0\\
-i\,\alpha_{+}(x)
\end{array}
\right),
\qquad
\chi_{{}_{\text{R}_{(2)}}}  &=e^{i\,\omega\,\varphi}
\left(
\begin{array}{@{}c@{}}
0\\
\alpha_{+}(x) \\
-i\,\alpha_{-}(x) \\
0
\end{array}
\right),
\\[2ex]
\chi_{{}_{\text{I}_{(1)}}} &  =e^{i\,\omega\,\varphi}
\left(
\begin{array}{@{}c@{}}
0\\
\alpha_{-}(x) \\
-i\,\alpha_{+}(x) \\
0
\end{array}
\right),
\qquad
\chi_{{}_{\text{I}_{(2)}}} &=e^{i\,\omega\,\varphi}\left(
\begin{array}{c}
-\alpha_{+}(x) \\
0\\
0\\
i\,\alpha_{-}(x)
\end{array}
\right),
\end{alignat}
where, from eq.\ \eqref{XIXR}, we have \,$\chi_{{}_{\text{I}_{(k)}}}\!=i\,\gamma^5\chi_{{}_{\text{R}_{(k)}}}$, and with
\begingroup
\setlength{\abovedisplayskip}{4pt plus 2pt minus 4pt}
\begin{equation}
\alpha_{\pm}(x) =\sqrt{\frac{\eta}{L}}\,\Upsilon(x)^{\frac{1}{4}}\,\sqrt{1\pm f(x)^{1/2}}\,,
\qquad\quad
\omega=-\frac{\pi}{\Delta}\,.\qquad
\end{equation}
\endgroup
The above relation involving $\Delta$ implies the antiperiodicity of the spinors. 
The chiral form of the spinors can be reproduced
as
\begin{equation}
\begin{split}
{\epsilon}_{(k)}^{1}=\operatorname{Re}\chi_{{}_{\text{R}_{(k)}}}+i\,\operatorname{Re}\chi_{{}_{\text{I}_{(k)}}}
=\left(\mathds{1}-\gamma^5\right)\,\operatorname{Re}\chi_{{}_{\text{R}_{(k)}}}\,,
\\[1.5ex]
{\epsilon}_{(k)}^{2}=\operatorname{Im}\chi_{{}_{\text{R}_{(k)}}}+i\,\operatorname{Im}\chi_{{}_{\text{I}_{(k)}}}
=\left(\mathds{1}-\gamma^5\right)\,\operatorname{Im}\chi_{{}_{\text{R}_{(k)}}}\,,
\end{split}
\end{equation}
giving
\begin{align}
{\epsilon}_{(1)}^{1}  &  =
\left(
\begin{array}[c]{r@{}l}
&\cos(\omega\varphi)\,\alpha_{-}(x) \\
i\,&\cos(\omega\varphi)\,\alpha_{-}(x)\\
i\,&\sin(\omega\varphi)\,\alpha_{+}(x)\\
&\sin(\omega\varphi)\,\alpha_{+}(x)
\end{array}
\right),
\qquad
{\epsilon}_{(1)}^{2}=
\left(
\begin{array}[c]{@{}r@{}l}
&\sin(\omega\varphi)\,\alpha_{-}(x)\\
i\,&\sin(\omega\varphi)\,\alpha_{-}(x)\\
-i\,&\cos(\omega\varphi)\,\alpha_{+}(x)\\
-&\cos(\omega\varphi)\,\alpha_{+}(x)
\end{array}
\right),\label{KS1}
\\[2.75ex]
{\epsilon}_{(2)}^{1}  &  =
\left(
\begin{array}{@{}r@{}l}
-i\,&\cos(\omega\varphi)\,\alpha_{+}(x)\\
&\cos(\omega\varphi)\,\alpha_{+}(x)\\
&\sin(\omega\varphi)\,\alpha_{-}(x)\\
-i\,&\sin(\omega\varphi)\,\alpha_{-}(x)
\end{array}
\right),
\qquad
{\epsilon}_{(2)}^{2}=
\left(
\begin{array}{@{}r@{}l}
-i\,&\sin(\omega\varphi)\,\alpha_{+}(x)\\
&\sin(\omega\varphi)\,\alpha_{+}(x)\\
-&\cos(\omega\varphi)\,\alpha_{-}(x)\\
i\,&\cos(\omega\varphi)\,\alpha_{-}(x)
\end{array}
\right),
\label{KS2}
\end{align}
that, as expected, satisfy the condition 
\begin{equation}
\gamma^{5}\,{\epsilon}_{(k)}^{A}=-{\epsilon}_{(k)}^{A}\:.
\end{equation}
If we consider the $\mathcal{N}=2$ framework, the existence of four chiral spinors identifies the solution as $1/2$ BPS \cite{Gallerati:2019mzs,Gallerati:2021cty}. 
The solution instead turns out to be $1/8$-BPS when referring to the maximal $\mathcal{N}=8$ theory.

\subsection{Supersymmetric solutions with fixed fluxes}
\label{subsec:susy_fixfl}
In order to satisfy the susy condition \eqref{eq:susyQ}, the fluxes must be related as (see eq.\ \eqref{eq:fluxes})
\begin{equation}
\psi_{2}=-\psi_{1}/x_{0}\:,
\end{equation}
that inserted in \eqref{fxp} determines the existence of supersymmetric solitons for:
\begin{alignat}{3}
&x_{0}=\frac{\psi_{1}}{1-\psi_{1}}\,,
\qquad
&&\psi_{2}=-1+\psi_{1}\,,
\qquad
&&\quad 0<\psi_{1}<1\,,
\label{bound}
\\[\jot]
&x_{0}=-\frac{\psi_{1}}{1+\psi_{1}}\,,
\qquad 
&&\psi_{2}=1+\psi_{1}\,,
\qquad
&& -1<\psi_{1}<0\,.
\end{alignat}
The EM configurations of \cite{Anabalon:2021tua} can be reproduced by taking $\psi_1 = \pm \frac{1}{2}$.

\subsection{Supersymmetric solutions with fixed charges}
\label{subsec:susy_fixch}
The susy condition in terms of the rescaled charges \eqref{eq:q1q2} simply reads
\begin{equation}
q_{1}=q_{2}\:.
\label{eq:susyq}
\end{equation}
The $\eta$ parameter \eqref{eq:eta_fixch} and the metric function \eqref{eq:fx0_fixch} now simplify in 
\begin{equation}
\eta=\frac{\Delta}{2\pi}\:\frac{1+x_0}{-1+x_0}\:,
\qquad\quad
f(x_0)=1-\frac{q_1^2\left(1+x_0\right)^4}{x_0}\:.\qquad
\end{equation}
Then, for every value \,$\left\vert q_{1}\right\vert<\tfrac{1}{4}$\, there exist two charged solitons (one for each branch \,$x_0\gtrless1$\,) satisfying the condition
\begin{equation}
f(x_0)=0
\qquad\rightarrow\qquad
q_{1}=\pm\frac{x_0}{(1+x_0)^2}\:,\qquad
\label{eq:q1x0}
\end{equation}
the sign depending on the sign of the charge.

\paragraph{Relation with pure EM solutions.}
It is possible to show that this choice of boundary conditions also features non-susy configuratons with vanishing scalar. 
%
If we consider the discussion of Sect.\ \ref{subsec:earlier}, it is easily found that EM solutions with vanishing scalar of \cite{Anabalon:2021tua} can be achieved from our susy configuration by the scaling $x_0\to1$ and 
\begin{equation}
q_1^2-q_2^2\to0\,.
\label{eq:q12scal}
\end{equation}
Since we have found susy hairy configuration \eqref{eq:susyq} satisfying the above condition but with $x_0\neq1$, we realize that in the fixed charges framework we can have both susy hairy and scalar-free EM solutions \cite{Anabalon:2021tua}, satisfying the same boundary condition \eqref{eq:q12scal}. The latter EM solutions however are \emph{not} supersymmetric (except when their period $\Delta$ is maximum) and correspond to the discussed additional intersection of the vertical dotted purple with the hairy solutions at $x_0=1$ in Fig.\ \ref{fig:sols_fixed_charges_1}.\par
When the period of the scalar-free solutions \cite{Anabalon:2021tua} takes the maximum value ${\Delta = \pi \sqrt{\frac{L^3}{Q}}}$, the configuration turns out to be supersymmetric, and coincides with our susy hairy soliton solution. In particular, since we have found the relation $Q = L^2\,\tilde Q_1$, this corresponds to a period for the hairy solution $\Delta = \pi \sqrt{\frac{L\,\eta}{Q_1}}$, that inserted in \eqref{eq:q1q2} gives
\begin{equation}
\Delta = \pi \sqrt{\tfrac{L\,\eta}{Q_1}}
\;\quad\Rightarrow\;\quad
q_1= \frac{1}{4}
\;\quad\Rightarrow\;\quad
x_0^2=1
\;\quad\Rightarrow\;\quad
\eta\to\infty
\end{equation}
having used equation \eqref{eq:q1x0}. This gives again the limit discussed in Sect.\ \ref{subsec:earlier}, where hairy configurations reduce to the vanishing scalar solutions \cite{Anabalon:2021tua}. The remarkable implications of this correspondence on the stability of susy configurations will be better illustrated in Sect.\ \ref{subsec:phase_fixch}.

\section{Phase structure}
\label{sec:phase}

\subsection{Euclidean action}
Let us consider the analytic continuation of the soliton metric to a real Euclidean metric $g_{\textsc{e}}$ featuring Euclidean time $\tau \in [0,\beta]$, \,$\beta=1/T$ representing the inverse of the temperature $T$ of the configuration. This would give rise to a periodic bosonic solution in $\beta$, being the gauge and the dilaton fields invariant under the continuation.

We note that, unless $\beta=\infty$, the thermal partition function is incompatible with supersymmetry, since it needs the fermions to be antiperiodic in $\tau$. As a result, only zero-temperature susy solutions exist, while non-susy bosonic solutions can be defined for all values of $\beta$.\par
The Euclidean action $S_{\textsc{e}}$ can be expressed as \cite{Anabalon:2020qux}:
\begin{equation}
\frac{S_{\textsc{e}}}{V}=I_{\text{bulk}}+I_{\textsc{gh}}+I_{\textsc{bk}}+I_{\text{ct}}+I_{\phi}\:,
\end{equation}
where
\begin{equation}
V=\beta\,\Delta\,\Delta_z\,,
\qquad\quad
\Delta_{z}=\int dz\,.    
\end{equation}
The $I_{\text{bulk}}$ term gives the bulk contribution,
\begin{equation}
I_{\text{bulk}}=\lim_{\epsilon\to 1^{-}}\;\frac{1}{\kappa}\:\int\limits_{x_{0}}^{\epsilon}dx\;\sqrt{g_{\textsc{e}}}\,\left(  -\frac{R}{2} + \frac{1}{2}\left(\partial\phi\right)^{2}  +\frac{1}{4}\,e^{\sqrt{2}\,\phi}\left(F^{1}\right)^{2}
+\frac{1}{4}e^{-\sqrt{2}\,\phi}\left(F^{2}\right)^{2}-\frac{2+\cosh\big(\sqrt{2}\,\phi\big)}{L^{2}}\right),
\end{equation}
while $I_{\textsc{gh}}$ is the Gibbons-Hawking term, $I_{\textsc{bk}}$ the Balasubramanian-Krauss counterterm, $I_{\text{ct}}$ is a scalar-dependent divergent counterterm and $I_{\phi}$ is a finite counterterm. For our solutions, they can be expressed as
\begingroup
\setlength{\belowdisplayskip}{1pt plus 2pt minus 4pt}
\begin{equation}
\begin{split}
I_{\textsc{gh}}=-\frac{1}{\kappa}\;\lim_{\epsilon\to1^{-}} K\,\sqrt{h}\,,
\qquad
I_{\textsc{bk}}=\frac{2}{\kappa\,L}\lim_{\epsilon\to1^{-}}\sqrt{h}\,,
\qquad
I_{\text{ct}}=\frac{1}{2\,\kappa\,L}\lim_{\epsilon\to1^{-}}\sqrt{h}\,\phi^{2}\,,
\qquad
I_{\phi}=0\,,
\end{split}
\end{equation}
\endgroup
where $h$ is the square root of the determinant of the boundary metric
$h_{\mu\nu}$ and $K$ is the trace of the extrinsic curvature $K_{\mu\nu}=\mathcal{D}_{\left(\mu\right.}N_{\left.\nu\right)}$.
The contributions explicitly read:
\begin{equation}
\begin{split}
I_{\text{bulk}} &  =\frac{1}{\kappa}\,\lim_{\rho\to\infty}\left(\frac
{\rho^{3}}{L^{4}}+\frac{\rho}{8\,\eta^{2}}-\frac{\mu}{L^{2}}\right),
\\[2ex]
I_{\textsc{gh}} &=-\frac{1}{\kappa}\,\lim_{\rho\to\infty}\left(\frac{3\,\rho^{3}}{L^{4}}+\frac{3\,\rho}{8\,\eta^{2}}-\frac{3\,\mu}{2\,L^{2}}\right),
\\[2ex]
I_{\textsc{bk}} & =\frac{1}{\kappa}\,\lim_{\rho\to\infty}\left(\frac{2\,\rho^{3}}{L^{4}}-\frac{\mu}{L^{2}}\right),
\\[2ex]
I_{\text{ct}}  &=\frac{1}{\kappa}\:\lim_{\rho\to\infty}\,\frac{\rho}{4\,\eta^{2}}\:,
\end{split}    
\end{equation}
giving for the Euclidean action
\begin{equation}
\frac{S_{\textsc{e}}}{V}=-\frac{\mu}{2\,L^{2}\,\kappa}\:,
\end{equation}
where $\mu$ given in \eqref{eq:mu}. Since there is no entropy associated with these configurations, their free energy corresponds to their energy.

\subsection{Fixed fluxes $\psi_1$ and $\psi_2$}
\begin{figure}[H]
\centering
\includegraphics[width=0.85\textwidth,keepaspectratio]{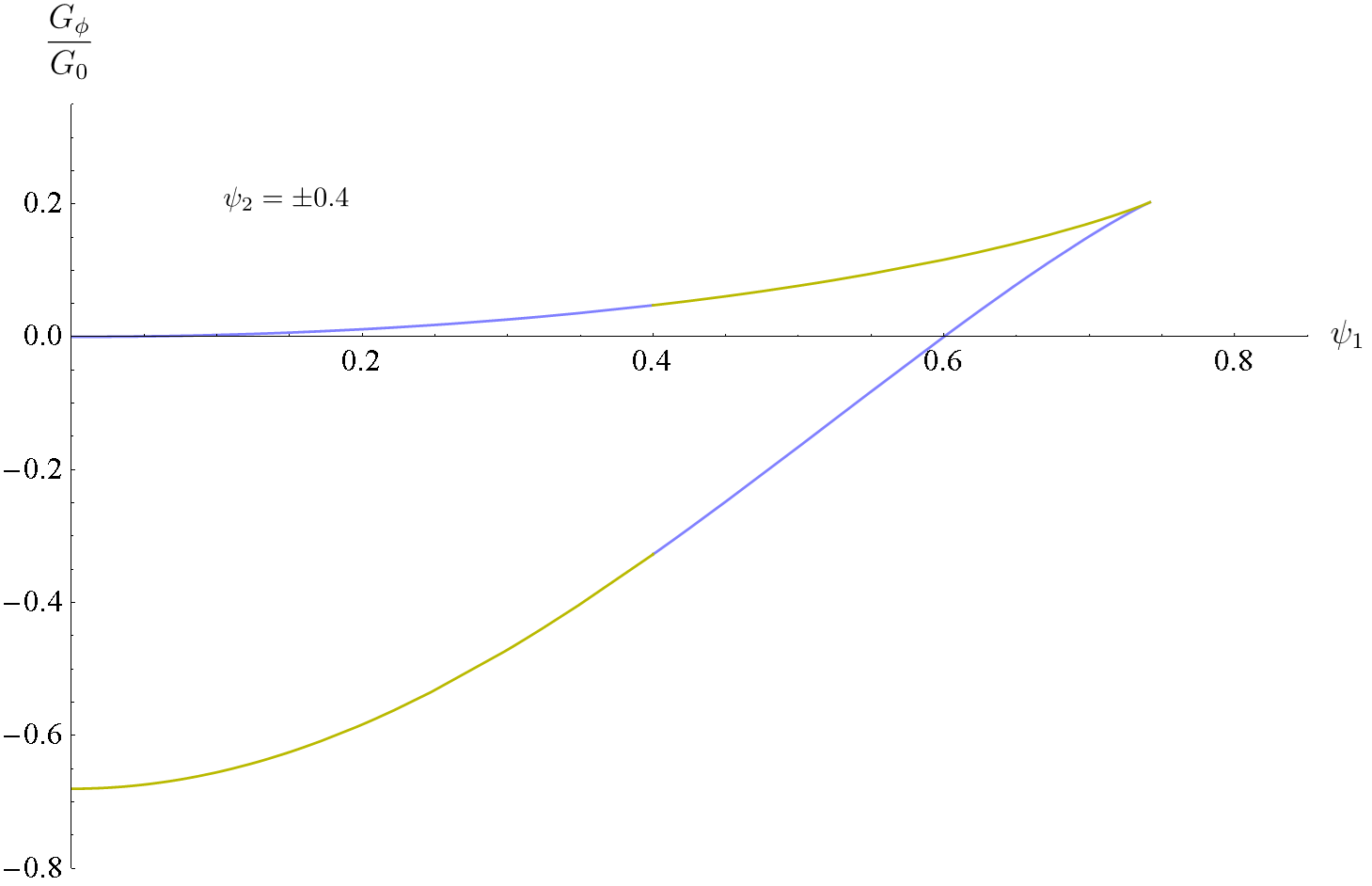}
\caption{Rescaled free energy $\frac{G_{\phi}}{\left\vert G_{0}\right\vert }$ as a function of $\psi_1$, for $\psi_{2}=\pm 0.4$. 
The $x\gtrless1$ branches of the solution are represented by different hues. The shift in color occurs when one of the spacetimes, pertaining to a configuration, transits into another (see Sect.\ \ref{sec:solutions}).}%
\label{fig:normGphi_fixed_fluxes_1}
\end{figure}

\begin{figure}[H]
\centering
\includegraphics[width=0.815\textwidth,keepaspectratio]{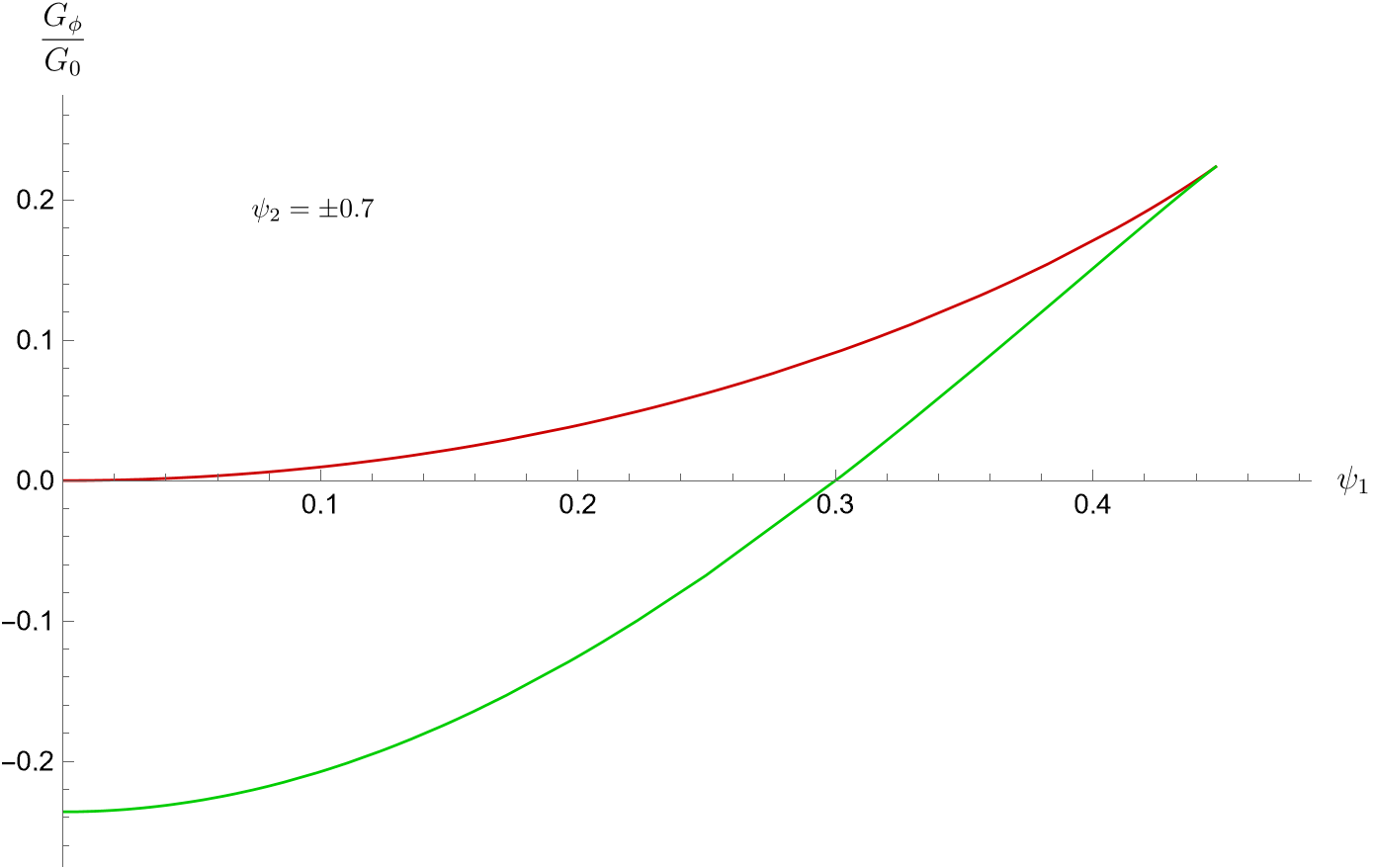}
\caption{Rescaled free energy $\frac{G_{\phi}}{\left\vert G_{0}\right\vert }$ as a function of $\psi_1$, for $\psi_{2}=\pm 0.7$. The $x\gtrless1$ branches of the solution are represented by different hues. The shift in color occurs when one of the spacetimes, pertaining to a configuration, transits into another (see Sect.\ \ref{sec:solutions}).}%
\label{fig:normGphi_fixed_fluxes_2}
\end{figure}

Let us now consider fixed fluxes boundary conditions. For the hairy soliton configuration, the free energy density is as follows:
\begin{equation}
\begin{split}
\frac{S_{\textsc{e}}}{V}  &  =\frac{G_{\phi}}{\Delta\,\Delta_{z}} =\frac{M}{\Delta\,\Delta_{z}}=
-\frac{\mu}{2\,\kappa\,L^2}
=\mp\frac{1}{2\,\kappa\,\eta}\left(Q_{1}^{2}-Q_{2}^{2}\right)=
\\[1ex]
& =\mp\frac{2\pi^{3}L^{2}}{\kappa\,\Delta^{3}}\:\frac{\left(  -\psi_{1}^{2}+x_{0}^{2}\,\psi_{2}^{2}\right)\left\vert(1+2\,x_0)\,\psi_{1}^{2}-x_0\,(2+x_0)\,\psi_{2}^{2}\right\vert}{x_0\left(-1+x_{0}^{2}\right)^{2}}\:,
\end{split}
\end{equation}
having employed eq.\ \eqref{Qfrompsi}. Using now the free energy of the AdS soliton of \cite{Horowitz:1998ha}
%
\begin{equation}
G_{0}=-\frac{32}{27}\:\frac{\pi^{3}L^{2}}{\Delta^{3}\,\kappa}\:\Delta\,\Delta_{z}\:.
\end{equation}
as a convenient normalization, we show in Figures \ref{fig:normGphi_fixed_fluxes_1}, \ref{fig:normGphi_fixed_fluxes_2} the ratio
\begin{equation}
\frac{G_{\phi}}{\left\vert G_{0}\right\vert }=\pm\frac{27}{16}\:\frac{\left(-\psi_{1}^{2}+x_{0}^{2}\,\psi_{2}^{2}\right)\left\vert(1+2\,x_0)\,\psi_{1}^{2}-x_0\,(2+x_0)\,\psi_{2}^{2}\right\vert
}{x_0\left(-1+x_{0}^{2}\right)^{2}}\:.
\end{equation}
We note that, due to the normalization $G_0$, the graphs always tend to a value larger than $-1$ in the vertical axis. This seems to suggest that the AdS soliton retains its status as the lowest energy configuration even within these more general supergravity theories, extending its significance beyond General Relativity. 

Figures \ref{fig:normGphi_fixed_fluxes_1} and \ref{fig:normGphi_fixed_fluxes_2} show two distinct branches of energies. The upper one ends up at zero energy, when there are no sources. The lower one ends up at a negative energy, and when there are no sources it matches the energy of the AdS Soliton. On the field theory side, the dual fermions are antiperiodic by construction on these solutions. Therefore, in the dual QFT, the higher energy branch corresponds to antiperiodic bosonic fields on the $\mathbb{S}^1$ and the zero energy end point of the higher energy branch is just a representation of the matching of the bosonic and fermionic degrees of freedom, as expected from a supersymmetric dual field theory. The lowest energy solution then has antiperiodic fermions and periodic bosons on the $\mathbb{S}^1$.
\par
\begin{figure}[H]
\centering
\includegraphics[width=0.85\textwidth,keepaspectratio]{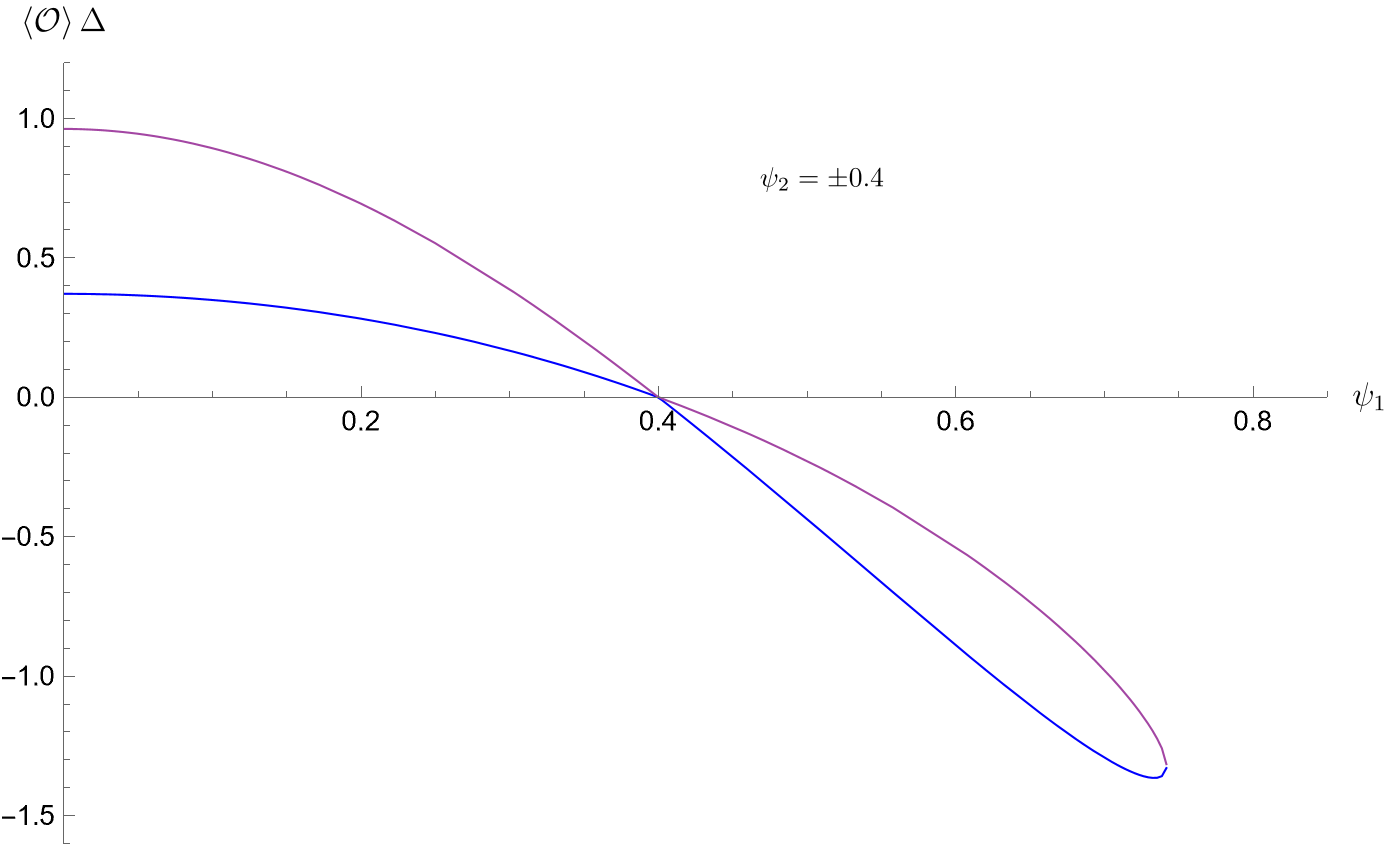}
\caption{Rescaled vev $\left\langle \mathcal{O}\right\rangle \Delta$ as a function of $\psi_1$, for $\psi_{2}=\pm 0.4$. Different
colours are used to represent different branches of the solution.}%
\label{fig:vevOphi_fixed_fluxes}
\end{figure}

\smallskip
It is remarkable that both the higher and lower energy solutions are reproduced in the gravity side by matching the negative vev and the positive vev solutions as can be seen from Figure \ref{fig:vevOphi_fixed_fluxes}, the latter representing the vev $\left\langle \mathcal{O}\right\rangle \Delta$ of eq.\ \eqref{eq:vev} originating from the presence of the dilaton.

Negative values of the vev $\left\langle \mathcal{O}\right\rangle \Delta$ in Figure \ref{fig:vevOphi_fixed_fluxes} correspond to configurations belonging to the $x>1$ branch, while the positive region is related to the $0<x<1$ interval. The susy solution (requiring $\mu=0$) is found for vanishing value of the free energy, therefore in correspondence with the intersection with the $\psi_{1}$ horizontal axis, see Sect.\ \ref{subsec:susy_fixfl}.

\subsection{Fixed charges $q_1$ and $q_2$}
\label{subsec:phase_fixch}
In the fixed charges boundary framework, the free energy of the solution is found by exploiting the Legendre transform of the Euclidean action 
\begin{equation}
\label{eq:Legendre}
\begin{split}
\frac{F_{\phi}}{V}&=\frac{S_{\textsc{e}}}{V}-\left.\left\langle J_\Lambda^{\nu}\right\rangle A^\Lambda_{\nu}\right|_{x\to1}=
\\[1ex]
&=-\frac{\mu}{2\,L^{2}\,\kappa}
-\langle J_1^{\varphi}\rangle\,
Q_1\left(1-x_0\right)
-\langle J_2^{\varphi}\rangle\, Q_2\left(-1+x_0^{-1}\right)=
\\[1ex]
&=\mp\frac{\left(Q_{1}^{2}-Q_{2}^{2}\right)}{2\,\eta\,\kappa}\pm\frac{Q_1^2}{\eta\,\kappa}\left(1-x_0\right)\pm\frac{Q_2^2}{\eta\,\kappa}\left(-1+x_0^{-1}\right)\,,
\end{split}    
\end{equation}
the signs depending on whether the solution falls within the range $x<1$ or $x>1$.

\paragraph{Supersymmetric solutions.}
In Sect.\ \ref{subsec:susy_fixch} we have seen how this fixed charges framework gives rise to the possibility of finding both susy hairy and scalar-free EM soliton configurations satisfying the same boundary conditions. This will lead to amazing implications on the (expected) stability of the supersymmetric configurations.
\par
Supersymmetric hairy solutions satisfy the condition \eqref{eq:susyq}, $q_1=q_2$\,. The rescaled charges \eqref{eq:q1q2} are related to the single charge of the EM configuration \cite{Anabalon:2021tua} as
\begin{equation}
q_1 =\frac{\Delta^{2}}{4\pi^{2}L}\:\frac{Q}{L^2}\:, \quad
\label{eq:q1Q}
\end{equation}
having used the relations of Sect.\ \ref{subsec:earlier}.\par\smallskip 
The free energy of the Einstein-Maxwell configuration \cite{Anabalon:2021tua} is given by
\begin{equation}
\begin{split}
\frac{F_{\textsc{em}}}{\Delta\,\Delta_z} =-\frac{\mu_0}{2\,\kappa\,L^{2}}+\frac{2\,Q^2}{\kappa\,L^2\,r_0}\,,
\qquad\quad 
\mu_0=\frac{r_0^4-L^2\,Q^2}{r_0\,L^2}\:,\qquad 
\end{split}
\end{equation}
with $\Delta=\frac{4\pi L^2\,r_0^3}{3\,r_0^4+L^2Q^2}$. Using relation \eqref{eq:q1Q}, we can rewrite the above EM free energy as 
\begin{equation}
\label{eq:EMfree}
\begin{split}
\frac{F_{\textsc{em}}}{\Delta\,\Delta_z} =\frac{2\pi^3L^2}{\Delta^3\,\kappa}\:X^2\,(5-4\,X)\:,
\end{split}
\end{equation}
with
\begingroup
\setlength{\belowdisplayshortskip}{14pt plus 2pt minus 2pt}
\setlength{\belowdisplayskip}{14pt plus 2pt minus 2pt}
\begin{equation}
q_1^2=\frac{1}{16}\,X^3\,(4-3\,X)\,,\qquad\quad
X=\frac{\Delta\,r_0}{\pi L^2}\,.\qquad 
\end{equation}
\endgroup
For the supersymmetric hairy solutions we find 
\begin{equation}
\frac{F_{\phi}}{|G_0|}=\frac{27}{4}\,|q_1|\:,  
\label{eq:hairyfree}
\end{equation}
having used \eqref{eq:q1q2}, \eqref{eq:eta_fixch}, \eqref{eq:susyq} and \eqref{eq:q1x0} in \eqref{eq:Legendre}.\par
\begin{figure}[!h]
\centering
\includegraphics[width=0.9\textwidth,keepaspectratio]{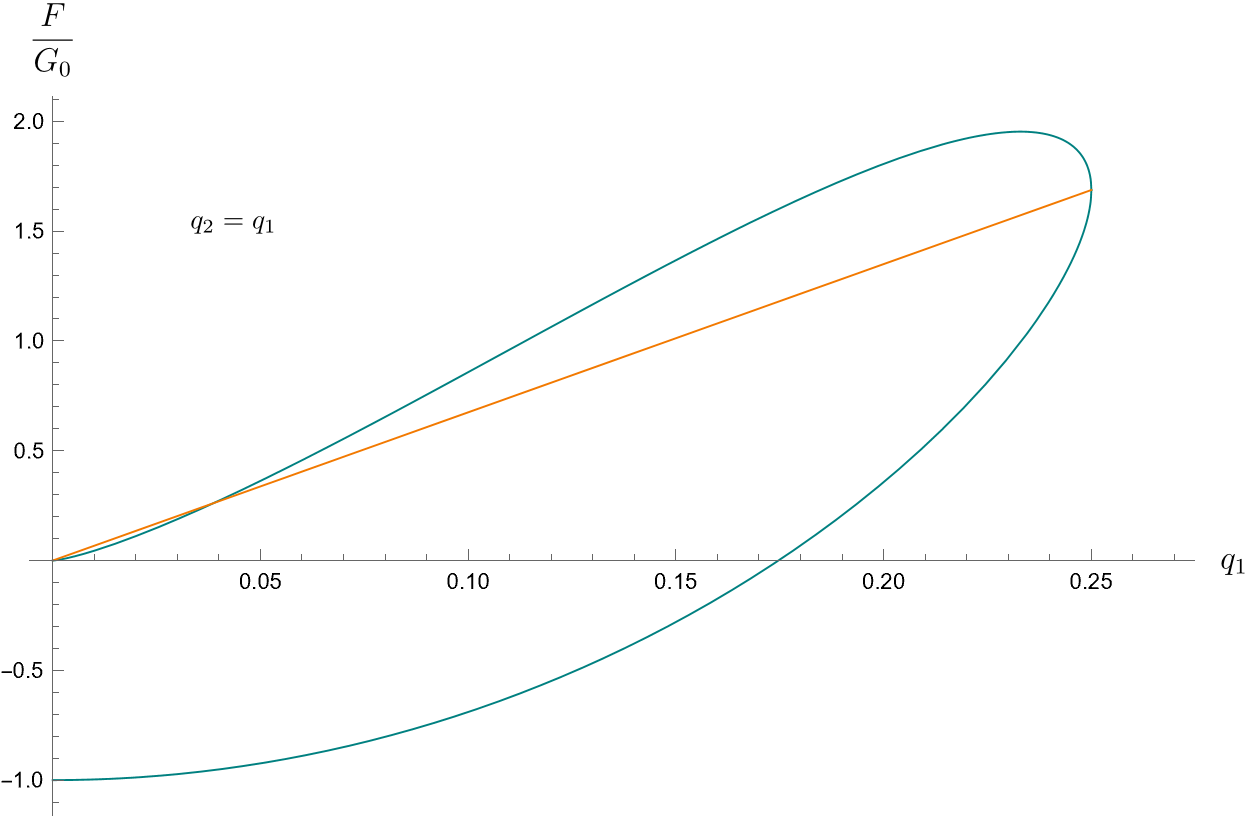}
\caption{Rescaled free energy $\frac{F}{\left\vert G_{0}\right\vert }$ as a function of $q_1$ for supersymmetric condition $q_1=q_2$. The orange line represents the hairy supersymmetric solitons \eqref{eq:hairyfree}, while the non-supersymmetric pure Einstein-Maxwell solitons \eqref{eq:EMfree} are shown in blue. }
\label{fig:normF_fixed _charges}%
\end{figure}

\smallskip
In order to compare the susy hairy solution and the non-susy EM configuration (featuring the same boundary conditions) we plot their free energies in Figure \ref{fig:normF_fixed _charges} as functions of $q_1$. We can see that both the Einstein-Maxwell and the hairy solitons exist for values of $q_1$ such that $q_1\leq\frac{1}{4}$. In the extreme value $q_1=\frac{1}{4}$ the different configurations merge yielding a supersymmetric soliton. 

The phase diagram illustrates the existence of a branch of non-susy Einstein-Maxwell solutions having lower free energy than the hairy supersymmetric solution. This may come as a surprise, as susy solutions satisfy a BPS bound that would be expected to identify them as the lowest energy configurations. 
This unconventional result, however, does not conflict with the positive energy theorem \cite{Witten:1981mf,Gibbons:1983aq}. The latter implies that the energy of a susy configuration is the lowest of the class featuring the same boundary conditions, but a necessary condition for the theorem to apply is the existence, for the non-susy configuration, of an asymptotic Killing spinor coinciding, up to $O(1/r^2)$ terms at infinity, with the Killing spinor of the susy one. The key point in the latter prescription lies in the definition of asymptotic Killing spinors arising in non-SUSY configurations. These spinors satisfy some Killing spinor equation only at the asymptotic region of the spacetime, not necessarily holding true for the entire space (as it happens for Killing spinors of a susy solutions). These properties are discussed more in detail in \cite{Anabalon:2023oge}.
\par
Since our susy hairy solutions have antiperiodic boundary conditions at infinity, the positive energy theorem applies only for non-susy solutions with an asymptotic Killing spinor with the same properties \cite{Anabalon:2022aig,Anabalon:2023oge}. This only happens for values of the charges at infinity where the susy solution's free energy is less than the non-supersymmetric one, as one would assume, preventing contradictions with the positive energy theorem implications.
The presented situation is then remarkable, as an example of framework in which an instability of the supersymmetric solutions under quantum phase transitions can occur \cite{Anabalon:2021smx}.

\bigskip
\bigskip

\section*{\normalsize Acknowledgements} 
We are grateful to Horatiu Nastase and Carlos N\'{u}ñez for interesting conversations. The work of AA, DA and JO is supported in part by the FONDECYT grants 1200986, 1210635, 1221504, 1230853 and 1242043. The work of AA is supported in part by the FAPESP visiting researcher award 2022/11765-7. DA acknowledges the hospitality of the Physics Department, Universidad de Concepción, during the last stages of this research. The work of D.A. is supported in part by the FONDECYT grant 1242043.


\bigskip\bigskip\bigskip

\appendix
\addtocontents{toc}{\protect\setcounter{tocdepth}{1}}
\addtocontents{toc}{\protect\addvspace{3.5pt}}%
\numberwithin{equation}{section}%
\numberwithin{figure}{section}%


\hypersetup{linkcolor=blue}
\phantomsection 
\addtocontents{toc}{\protect\addvspace{4.5pt}}
\addcontentsline{toc}{section}{References} 
\bibliographystyle{mybibstyle}
\bibliography{bibliografia}

\providecommand{\href}[2]{#2}\begingroup\begin{thebibliography}{10}

\bibitem{Maldacena:1997re}
Juan~Martin Maldacena, \textit{``{The Large N limit of superconformal field
  theories and supergravity}''}, Adv. Theor. Math. Phys. \textbf{2} (1998)
  231--252,
  [\href{http://arxiv.org/abs/hep-th/9711200}{\texttt{hep-th/9711200}}].

\bibitem{Witten:1998zw}
Edward Witten, \textit{``{Anti-de Sitter space, thermal phase transition, and
  confinement in gauge theories}''}, Adv. Theor. Math. Phys. \textbf{2} (1998)
  505--532,
  [\href{http://arxiv.org/abs/hep-th/9803131}{\texttt{hep-th/9803131}}].

\bibitem{Horowitz:1998ha}
Gary~T. Horowitz and Robert~C. Myers, \textit{``{The AdS / CFT correspondence
  and a new positive energy conjecture for general relativity}''}, Phys. Rev. D
  \textbf{59} (1998) 026005,
  [\href{http://arxiv.org/abs/hep-th/9808079}{\texttt{hep-th/9808079}}].

\bibitem{Witten:1981gj}
Edward Witten, \textit{``{Instability of the Kaluza-Klein Vacuum}''}, Nucl.
  Phys. B \textbf{195} (1982) 481--492.

\bibitem{Anabalon:2021tua}
Andres Anabalon and Simon~F. Ross, \textit{``{Supersymmetric solitons and a
  degeneracy of solutions in AdS/CFT}''}, JHEP \textbf{07} (2021) 015,
  [\href{http://arxiv.org/abs/2104.14572}{\texttt{arXiv:2104.14572}}].

\bibitem{Anabalon:2023oge}
Andres Anabalon, Mattia Cesaro, Antonio Gallerati, Alfredo Giambrone and Mario
  Trigiante, \textit{``{A positive energy theorem for AdS solitons}''}, Phys.
  Lett. B \textbf{846} (2023) 138226,
  [\href{http://arxiv.org/abs/2304.09201}{\texttt{arXiv:2304.09201}}].

\bibitem{Anabalon:2022aig}
Andr\'es Anabal\'on, Antonio Gallerati, Simon Ross and Mario Trigiante,
  \textit{``{Supersymmetric solitons in gauged $\mathcal{N}=8$
  supergravity}''},
  \href{http://arxiv.org/abs/2210.06319}{\texttt{arXiv:2210.06319}}.

\bibitem{Canfora:2021nca}
Fabrizio Canfora, Julio Oliva and Marcelo Oyarzo, \textit{``{New BPS solitons
  in $ \mathcal{N} $ = 4 gauged supergravity and black holes in
  Einstein-Yang-Mills-dilaton theory}''}, JHEP \textbf{02} (2022) 057,
  [\href{http://arxiv.org/abs/2111.11915}{\texttt{arXiv:2111.11915}}].

\bibitem{Nunez:2023nnl}
Carlos Nunez, Marcelo Oyarzo and Ricardo Stuardo, \textit{``{Confinement in (1
  + 1) dimensions: a holographic perspective from I-branes}''}, JHEP
  \textbf{09} (2023) 201,
  [\href{http://arxiv.org/abs/2307.04783}{\texttt{arXiv:2307.04783}}].

\bibitem{Nunez:2023xgl}
Carlos Nunez, Marcelo Oyarzo and Ricardo Stuardo, \textit{``{Confinement and D5
  branes}''},
  \href{http://arxiv.org/abs/2311.17998}{\texttt{arXiv:2311.17998}}.

\bibitem{Fatemiabhari:2024aua}
Ali Fatemiabhari and Carlos Nunez, \textit{``{From conformal to confining field
  theories using holography}''},
  \href{http://arxiv.org/abs/2401.04158}{\texttt{arXiv:2401.04158}}.

\bibitem{Anabalon:2022ksf}
Andr\'es Anabal\'on, Patrick Concha, Julio Oliva, Constanza Quijada and Evelyn
  Rodr\'\i{}guez, \textit{``{Phase transitions for charged planar solitons in
  AdS}''}, \href{http://arxiv.org/abs/2205.01609}{\texttt{arXiv:2205.01609}}.

\bibitem{Quijada:2023fkc}
Constanza Quijada, Andr\'es Anabal\'on, Robert~B. Mann and Julio Oliva,
  \textit{``{Triple Points of Gravitational AdS Solitons and Black Holes}''},
  \href{http://arxiv.org/abs/2308.16341}{\texttt{arXiv:2308.16341}}.

\bibitem{Durgut:2023rmu}
Turkuler Durgut and Hari~K. Kunduri, \textit{``{Supersymmetric asymptotically
  locally AdS5 gravitational solitons}''}, Annals Phys. \textbf{457} (2023)
  169435, [\href{http://arxiv.org/abs/2307.02466}{\texttt{arXiv:2307.02466}}].

\bibitem{Cvetic:1999xx}
Mirjam Cvetic, S.~S. Gubser, Hong Lu and C.~N. Pope, \textit{``{Symmetric
  potentials of gauged supergravities in diverse dimensions and Coulomb branch
  of gauge theories}''}, Phys. Rev. D \textbf{62} (2000) 086003,
  [\href{http://arxiv.org/abs/hep-th/9909121}{\texttt{hep-th/9909121}}].

\bibitem{Cvetic:1999xp}
Mirjam Cvetic, M.~J. Duff, P.~Hoxha, James~T. Liu, Hong Lu, J.~X. Lu,
  R.~Martinez-Acosta, C.~N. Pope, H.~Sati and Tuan~A. Tran,
  \textit{``{Embedding AdS black holes in ten-dimensions and
  eleven-dimensions}''}, Nucl. Phys. B \textbf{558} (1999) 96--126,
  [\href{http://arxiv.org/abs/hep-th/9903214}{\texttt{hep-th/9903214}}].

\bibitem{Luciani:1977hp}
J.~F. Luciani, \textit{``{Coupling of O(2) Supergravity with Several Vector
  Multiplets}''}, Nucl. Phys. B \textbf{132} (1978) 325--332.

\bibitem{Gallerati:2016oyo}
Antonio Gallerati and Mario Trigiante, \textit{``{Introductory Lectures on
  Extended Supergravities and Gaugings}''}, Springer Proc. Phys. \textbf{176}
  (2016) 41--109,
  [\href{http://arxiv.org/abs/1809.10647}{\texttt{arXiv:1809.10647}}].

\bibitem{Lauria:2020rhc}
Edoardo Lauria and Antoine Van~Proeyen, \textit{``{${\cal N}=2$ Supergravity in
  $D=4,5,6$ Dimensions}''}; vol.~966 (3, 2020).

\bibitem{Anabalon:2020pez}
Andres Anabalon, Dumitru Astefanesei, Antonio Gallerati and Mario Trigiante,
  \textit{``{New non-extremal and BPS hairy black holes in gauged
  $\,\mathcal{N}=2\,$ and $\,\mathcal{N}=8\,$ supergravity}''}, JHEP
  \textbf{04} (2021) 047,
  [\href{http://arxiv.org/abs/2012.09877}{\texttt{arXiv:2012.09877}}].

\bibitem{Anabalon:2021smx}
Andr\'es Anabal\'on, Dumitru Astefanesei, Antonio Gallerati and Mario
  Trigiante, \textit{``{Instability of supersymmetric black holes via quantum
  phase transitions}''}, JHEP \textbf{11} (2021) 116,
  [\href{http://arxiv.org/abs/2105.08771}{\texttt{arXiv:2105.08771}}].

\bibitem{Anabalon:2012ta}
Andres Anabalon, \textit{``{Exact Black Holes and Universality in the
  Backreaction of non-linear Sigma Models with a potential in (A)dS4}''}, JHEP
  \textbf{06} (2012) 127,
  [\href{http://arxiv.org/abs/1204.2720}{\texttt{arXiv:1204.2720}}].

\bibitem{Anabalon:2019tcy}
Andres Anabalon, Dumitru Astefanesei, David Choque and Jose~D. Edelstein,
  \textit{``{Phase transitions of neutral planar hairy AdS black holes}''},
  JHEP \textbf{07} (2020) 129,
  [\href{http://arxiv.org/abs/1912.03318}{\texttt{arXiv:1912.03318}}].

\bibitem{Anabalon:2017yhv}
Andr\'es Anabal\'on, Dumitru Astefanesei, Antonio Gallerati and Mario
  Trigiante, \textit{``{Hairy Black Holes and Duality in an Extended
  Supergravity Model}''}, JHEP \textbf{04} (2018) 058,
  [\href{http://arxiv.org/abs/1712.06971}{\texttt{arXiv:1712.06971}}].

\bibitem{Duff:1995sm}
M.~J. Duff, James~T. Liu and J.~Rahmfeld, \textit{``{Four-dimensional
  string-string-string triality}''}, Nucl. Phys. B \textbf{459} (1996)
  125--159,
  [\href{http://arxiv.org/abs/hep-th/9508094}{\texttt{hep-th/9508094}}].

\bibitem{Behrndt:1996hu}
Klaus Behrndt, Renata Kallosh, Joachim Rahmfeld, Marina Shmakova and Wing~Kai
  Wong, \textit{``{STU black holes and string triality}''}, Phys. Rev. D
  \textbf{54} (1996) 6293--6301,
  [\href{http://arxiv.org/abs/hep-th/9608059}{\texttt{hep-th/9608059}}].

\bibitem{Behrndt:1997ny}
Klaus Behrndt, Dieter Lust and Wafic~A. Sabra, \textit{``{Stationary solutions
  of N=2 supergravity}''}, Nucl. Phys. B \textbf{510} (1998) 264--288,
  [\href{http://arxiv.org/abs/hep-th/9705169}{\texttt{hep-th/9705169}}].

\bibitem{Duff:1999gh}
M.~J. Duff and James~T. Liu, \textit{``{Anti-de Sitter black holes in gauged N
  = 8 supergravity}''}, Nucl. Phys. B \textbf{554} (1999) 237--253,
  [\href{http://arxiv.org/abs/hep-th/9901149}{\texttt{hep-th/9901149}}].

\bibitem{Andrianopoli:2013kya}
Laura Andrianopoli, Riccardo D'Auria, Antonio Gallerati and Mario Trigiante,
  \textit{``{Extremal Limits of Rotating Black Holes}''}, JHEP \textbf{05}
  (2013) 071,
  [\href{http://arxiv.org/abs/1303.1756}{\texttt{arXiv:1303.1756}}].

\bibitem{Andrianopoli:2013jra}
Laura Andrianopoli, Antonio Gallerati and Mario Trigiante, \textit{``{On
  Extremal Limits and Duality Orbits of Stationary Black Holes}''}, JHEP
  \textbf{01} (2014) 053,
  [\href{http://arxiv.org/abs/1310.7886}{\texttt{arXiv:1310.7886}}].

\bibitem{Andrianopoli:2013ksa}
L.~Andrianopoli, R.~D'Auria, A.~Gallerati and M.~Trigiante, \textit{``{On $D =
  4$ Stationary Black Holes}''}, J. Phys. Conf. Ser. \textbf{474} (2013)
  012002.

\bibitem{Myers:1999psa}
Robert~C. Myers, \textit{``{Stress tensors and Casimir energies in the AdS /
  CFT correspondence}''}, Phys. Rev. D \textbf{60} (1999) 046002,
  [\href{http://arxiv.org/abs/hep-th/9903203}{\texttt{hep-th/9903203}}].

\bibitem{Anabalon:2015xvl}
Andres Anabalon, Dumitru Astefanesei, David Choque and Cristian Martinez,
  \textit{``{Trace Anomaly and Counterterms in Designer Gravity}''}, JHEP
  \textbf{03} (2016) 117,
  [\href{http://arxiv.org/abs/1511.08759}{\texttt{arXiv:1511.08759}}].

\bibitem{Astefanesei:2018vga}
Dumitru Astefanesei, Romina Ballesteros, David Choque and Ra\'ul Rojas,
  \textit{``{Scalar charges and the first law of black hole thermodynamics}''},
  Phys. Lett. B \textbf{782} (2018) 47--54,
  [\href{http://arxiv.org/abs/1803.11317}{\texttt{arXiv:1803.11317}}].

\bibitem{Astefanesei:2021ryn}
Dumitru Astefanesei, David Choque, Jorge Maggiolo and Ra\'ul Rojas,
  \textit{``{Holography of AdS hairy black holes and Cardy-Verlinde
  formula}''}, Phys. Rev. D \textbf{106} (2022), n.~4, 044032,
  [\href{http://arxiv.org/abs/2111.01337}{\texttt{arXiv:2111.01337}}].

\bibitem{Marolf:2006nd}
Donald Marolf and Simon~F. Ross, \textit{``{Boundary Conditions and New
  Dualities: Vector Fields in AdS/CFT}''}, JHEP \textbf{11} (2006) 085,
  [\href{http://arxiv.org/abs/hep-th/0606113}{\texttt{hep-th/0606113}}].

\bibitem{Kastor:2020wsm}
David Kastor and Jennie Traschen, \textit{``{Geometry of AdS-Melvin
  Spacetimes}''}, Class. Quant. Grav. \textbf{38} (2021), n.~4, 045016,
  [\href{http://arxiv.org/abs/2009.14771}{\texttt{arXiv:2009.14771}}].

\bibitem{Gallerati:2019mzs}
Antonio Gallerati, \textit{``{Constructing black hole solutions in supergravity
  theories}''}, Int. J. Mod. Phys. A \textbf{34} (2020), n.~35, 1930017,
  [\href{http://arxiv.org/abs/1905.04104}{\texttt{arXiv:1905.04104}}].

\bibitem{Gallerati:2021cty}
Antonio Gallerati, \textit{``{New Black Hole Solutions in $N = 2$ and $N = 8$
  Gauged Supergravity}''}, Universe \textbf{7} (2021), n.~6, 187.

\bibitem{Anabalon:2020qux}
A.~Anabal\'on, D.~Astefanesei, D.~Choque, A.~Gallerati and M.~Trigiante,
  \textit{``{Exact holographic RG flows in extended SUGRA}''}, JHEP \textbf{04}
  (2021) 053,
  [\href{http://arxiv.org/abs/2012.01289}{\texttt{arXiv:2012.01289}}].

\bibitem{Witten:1981mf}
Edward Witten, \textit{``{A Simple Proof of the Positive Energy Theorem}''},
  Commun. Math. Phys. \textbf{80} (1981) 381.

\bibitem{Gibbons:1983aq}
G.~W. Gibbons, C.~M. Hull and N.~P. Warner, \textit{``{The Stability of Gauged
  Supergravity}''}, Nucl. Phys. B \textbf{218} (1983) 173.

\end{thebibliography}\endgroup

\end{document}